\documentclass[traditabstract]{aa}

\usepackage{natbib}
\usepackage{graphicx}
\usepackage[bookmarks=false, colorlinks=true, citecolor=blue, linkcolor=blue]{hyperref}

\def\micron{\hbox{\,$\mu$m}}
\newcommand{\Lsun}{\hbox{$L_{\rm \odot}$}}
\newcommand{\Msun}{\hbox{$M_{\rm \odot}$}}
\newcommand{\degree}{\ensuremath{^\circ}}

\bibpunct{(}{)}{;}{a}{}{,}

\titlerunning{Sub-kpc SF law in the LIRG IC~4687}
\authorrunning{Pereira-Santaella et al.}

\begin{document}

\title{Sub-kpc star formation law in the local luminous infrared galaxy IC~4687 as seen by ALMA}
\author{M. Pereira-Santaella\inst{1, 2}, L. Colina\inst{1, 2}, S. Garc\'ia-Burillo\inst{3}, P. Planesas\inst{3}, A. Usero\inst{3}, A. Alonso-Herrero\inst{4}, S. Arribas\inst{1, 2}, S. Cazzoli\inst{1, 2}, B. Emonts\inst{1, 2}, J. Piqueras L\'opez\inst{1, 2}, and M. Villar-Mart\'in\inst{1, 2} 
}
\institute{Centro de Astrobiolog\'ia (CSIC/INTA), Ctra de Torrej\'on a Ajalvir, km 4, 28850, Torrej\'on de Ardoz, Madrid, Spain\\ \email{mpereira@cab.inta-csic.es}\label{inst1},
ASTRO-UAM, UAM, Unidad Asociada CSIC\label{inst2},
Observatorio Astron\'omico Nacional (OAN-IGN)-Observatorio de Madrid, Alfonso XII, 3, 28014, Madrid, Spain\label{inst3} ,
Instituto de F\'isica de Cantabria, CSIC-Universidad de Cantabria, E-39005 Santander, Spain\label{inst4}
}

\abstract{We analyze the spatially resolved (250\,pc scales) and integrated star formation (SF) law in the local luminous infrared galaxy (LIRG) IC~4687. This is one of the first studies of the SF law on a starburst LIRG at these small spatial scales.
We combined new interferometric ALMA CO(2--1) data with existing \textit{HST}\slash NICMOS Pa$\alpha$ narrowband imaging and VLT\slash SINFONI near-IR integral field spectroscopy to obtain accurate extinction-corrected SF rate (SFR) and cold molecular gas surface densities ($\Sigma_{\rm gas}$ and $\Sigma_{\rm SFR}$).
We find that IC~4687 forms stars very efficiently with an average depletion time ($t_{\rm dep}$) of 160\,Myr for the individual 250\,pc regions. This is approximately one order of magnitude shorter than the $t_{\rm dep}$ of local normal spirals and also shorter than that of main-sequence high-$z$ objects, even when we use a Galactic $\alpha_{\rm CO}$ conversion factor. This result suggests a bimodal SF law in the $\Sigma_{\rm SFR} \propto \Sigma_{\rm gas}^{N}$ representation. A universal SF law is recovered if we normalize the $\Sigma_{\rm gas}$ by the global dynamical time. However, at the spatial scales studied here, we find that the SF efficiency (or $t_{\rm dep}$) does not depend on the local dynamical time for this object. Therefore, an alternative normalization (e.g., free-fall time) should be found if a universal SF law exists at these scales.
}


\maketitle

\section{Introduction}\label{s_intro}

There is a strong correlation between the star formation rate (SFR) and the cold molecular gas content in galaxies. This relation is usually referred to as the star formation (SF) law (or as the Kennicutt-Schmidt relation; \citealt{Schmidt1959, Kennicutt1998b}) and it is expressed as
\begin{equation}
\Sigma_{\rm SFR} = A \Sigma_{\rm gas}^{N}
,\end{equation}
where $\Sigma_{\rm SFR}$ and $\Sigma_{\rm gas}$ are the SFR and cold molecular gas surface densities, respectively. For galaxy integrated observations, the typical power-law index, $N$, is 1.4--1.5 \citep{Kennicutt1998b, Yao2003}. The physical processes leading to this value of $N$ are not well established yet, although theoretical models suggest that variations of the free-fall time ($t_{\rm ff}$) and the orbital dynamical time might define the observed relation (see \citealt{McKee2007} and \citealt{Kennicutt2012} for a review).

In general, it is assumed that the normalization of the SF law, $A$, is constant, that is, independent of the galaxy type. However, some works have found bimodal SF laws when main sequence (MS) and starbursts (those with higher specific SFR than MS galaxies for a given redshift) are considered (e.g., \citealt{Daddi2010b,Genzel2010,GarciaBurillo2012}). In these cases, normal galaxies have depletion times ($t_{\rm dep} = M_{\rm H_2}\slash {\rm SFR}$) between 4 and 10 times longer than starburst galaxies. The possible existence of this bimodality in the SF law affects the determination of $N$. Actually, these works find an almost linear relation, $N\sim1$, for each galaxy population (MS and starbursts) when they are treated independently.

Recently, many studies of the resolved sub-kpc SF laws in nearby galaxies have appeared (e.g., \citealt{Kennicutt2007, Bigiel2008, Leroy2008, Verley2010, Rahman2012, Viaene2014, Casasola2015}). Most of these works find a wide range of $N$ values (0.8--2.3) and a considerable scatter in the relations (0.1--0.4\,dex). This could be explained if the SF law breaks down on sub-kpc scales (e.g., the location of the cold molecular gas peaks, CO, and the SFR regions, H$\alpha$, are not always coincident; \citealt{Kennicutt2007, Schruba2010}) and\slash or because some systematics affect these sub-kpc studies (e.g., the treatment of the diffuse background emission; \citealt{Liu2011}). 

\begin{figure*}[th!]
\centering
\includegraphics[width=\textwidth]{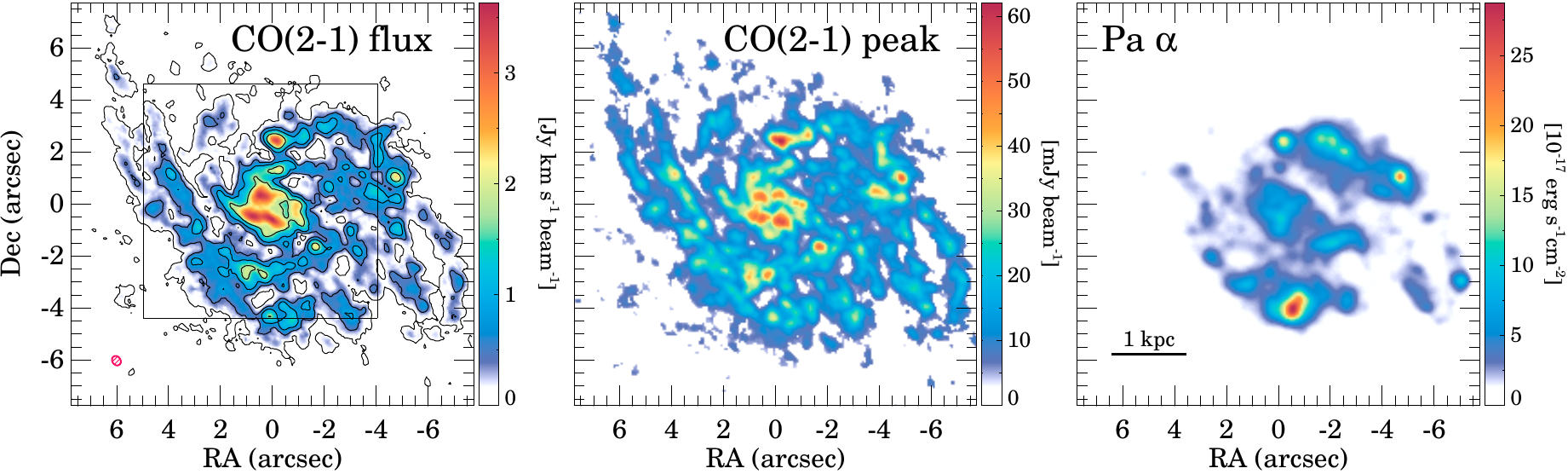}
\caption{ALMA $^{12}$CO(2--1) integrated flux and peak intensity for $\sim$5\,km\,s$^{-1}$ channels (left and middle panels) and \textit{HST}\slash NICMOS Pa$\alpha$ map smoothed to the ALMA resolution (right panel) of IC~4687. The contour levels of the left panel indicate the 10, 100, 200, and 400$\sigma$ levels. The 1$\sigma$ sensitivity of the data is 1\,mJy\,beam$^{-1}$ for $\sim$5\,km\,s$^{-1}$ channels.
The red hatched ellipse in the left panel indicates the ALMA beam size (0\farcs31$\times$0\farcs39, PA 35\degree). The black square indicates the field of view covered by the VLT\slash SINFONI data, i.e., the region to which we restricted our analysis in this paper. 
\label{fig_maps}}
\end{figure*}

These previous sub-kpc studies are focused on very nearby ($d<20$\,Mpc) spiral galaxies and active galactic nuclei (AGN). That is, the only objects where sub-kpc resolutions could be achieved before the arrival of the Atacama Large Millimeter/submillimeter Array (ALMA). Therefore, the most extreme local starbursts (i.e., luminous and ultraluminous IR galaxies) are absent in previous sub-kpc studies, although they are important to understand extreme high-$z$ SF (e.g., \citealt{Daddi2010b}).

In this paper, we present one of the first sub-kpc analyses of the SF law in a local extreme starburst. In particular, we study the local ($d$=74\,Mpc; 345\,pc\,arcsec$^{-1}$) luminous IR galaxy (LIRG) IC~4687. This galaxy has an IR luminosity of 10$^{11.3}$\Lsun, which corresponds to an integrated SFR of $\sim$30\,\Msun\,yr$^{-1}$ \citep{Pereira2011}. Although, the energy output of IC~4687 is dominated by SF, a weak AGN with a contribution $<$5\,\%\ to the total IR luminosity, is possibly present \citep{AAH2012a}. IC~4687 forms part of an interacting group together with IC~4686 and IC~4689, which are $\sim$10 and $\sim$20\,kpc away from IC~4687, respectively. Both IC~4687 and IC~4689 are spiral-like galaxies with ordered velocity fields that are dominated by rotation and kinematic centers coincident with their optical continuum peaks. Only the less massive galaxy of the system (IC~4686) shows a velocity field dominated by a tidal tail (see \citealt{Bellocchi2013} for details). Therefore, the starburst of IC~4687 might be induced by this weak interaction.

We obtained new ALMA $^{12}$CO(2--1) observations with $\sim$100 pc spatial resolution to study the SF law in IC~4687. We combined these observations with \textit{HST}\slash NICMOS maps of Pa$\alpha$ (50\,pc resolution; \citealt{AAH06s}) and VLT\slash SINFONI near-IR integral field spectroscopy (200\,pc resolution; \citealt{Piqueras2012, Piqueras2013, Piqueras2015}). This multi-wavelength dataset allowed us to get a novel insight into the sub-kpc SF law in extreme local starbursts.
In addition, optical integral field spectroscopy data of the entire IC~4686/4687/4689 system have recently been  obtained \citep{RodriguezZaurin2011, Bellocchi2013, Arribas2014, Cazzoli2015}, but they were not considered in the present analysis because of their different spatial resolution.

This paper is organized as follows: We describe the observations and data reduction in Section \ref{s_data}. The analysis of the cold molecular gas and SFR maps of IC~4687 are presented in Section \ref{s_analysis}. In Sections \ref{s_local} and \ref{s_higz}, we discuss our results in the context of nearby and high-$z$ galaxies, respectively. Finally, in Section \ref{s_conclusions}, we summarize the main findings of the paper.

\section{Observations and data reduction}\label{s_data}

\subsection{$^{12}$CO(2--1) ALMA data}\label{s_co_alma}

We obtained band 6 observations of IC~4687 with ALMA on August 28 2014 and April 5 2015 using extended and compact antenna array configurations with 35 and 39 antennas, respectively (project 2013.1.00271.S; PI: L. Colina). The on-source integration times were 18 and 9\,min, respectively. Both observations were single pointing centered at the nucleus of IC~4687. The extended configuration had baselines between 33.7\,m and 1.1\,km, while  the baselines ranged between 15.1\,m and 328\,m for the compact configuration. For these baselines, the maximum recoverable scales are 4.9\arcsec\ and 10.9\arcsec, respectively.

Two spectral windows of 1.875\,GHz bandwidth (0.48\,MHz~$\sim$ 0.6\,km\slash s channels) were centered at the sky frequencies of $^{12}$CO(2--1) (226.4\,GHz) and CS(5--4) (240.7\,GHz). In addition, two continuum spectral windows were set at 228.6 and 243.4\,GHz. In this paper, we only present the analysis of the CO(2--1) data.

The two datasets were calibrated using the standard ALMA reduction software CASA (v4.2.2; \citealt{McMullin2007}).  We used J1617-5848 for the amplitude calibration, assuming a flux density of 0.651\,Jy at 226.4\,GHz, and Titan, using the Butler-JPL-Horizons 2012 model, for the extended and compact configurations, respectively.
The $uv$ visibilities of each observation were converted to a common frequency reference frame (kinematic local standard of rest; LRSK) and then combined. The amplitudes of the baselines in common for both array configurations were in good agreement.
Then, the continuum (0.15-0.05\,mJy\,beam$^{-1}$) was fitted with the line free channels and subtracted in the $uv$ plane. In the final data cubes, we used 4\,MHz channels ($\sim$5\,km\,s$^{-1}$) and 256$\times$256 pixels of 0\farcs07. For the cleaning, we used the Briggs weighting with a robustness parameter of 0.5 \citep{Briggs1995PhDT}, which provided a beam with a full width half maximum (FWHM) of 0\farcs31$\times$0\farcs39 ($\sim$100\,pc$\times$130\,pc) with a position angle (PA) of 35\degree.
A mask derived from the observed CO(2--1) emission in each channel was used during the clean process. For the final 4\,MHz channels, the achieved 1$\sigma$ sensitivity is $\sim$1\,mJy\,beam$^{-1}$.
We applied the primary beam correction to the data cubes.

The integrated CO(2--1) flux in the considered ALMA field of view (18\arcsec$\times$18\arcsec) is 460\,Jy\,km\,s$^{-1}$ with a flux calibration uncertainty about 15\%. For comparison, the single-dish CO(2--1) flux measured with the 15\,m SEST telescope (24\arcsec\ beam size) is 480\,Jy\,km\,s$^{-1}$ \citep{Albrecht2007}. Therefore, combining both the compact and the extended ALMA array configurations we are able to recover most of the CO(2--1) flux of this source.

\subsection{Ancillary \textit{HST}\slash NICMOS and \textit{VLT}\slash SINFONI data}

We used the continuum subtracted narrowband Pa$\alpha$ \textit{HST}\slash NICMOS image of IC~4687 \citep{AAH06s} to determine the resolved SFR. The original Pa$\alpha$ map (0\farcs15 resolution) was convolved with a Gaussian kernel to match the angular resolution of the ALMA map.
To correct the Pa$\alpha$ emission for extinction (see next section), we used the 2D extinction maps of this galaxy derived with near-IR VLT\slash SINFONI integral field spectroscopy \citep{Piqueras2013, Piqueras2015}.
Both the ALMA CO(2--1) and the NICMOS Pa$\alpha$ maps cover similar fields of view. However, we limited our analysis to the smaller field of view of the SINFONI extinction map (8\arcsec$\times$8\arcsec; see Figure \ref{fig_maps}) so the dataset  would be homogeneously corrected for extinction. Nevertheless, this region contains $\sim$85\%\ of the total CO(2--1) flux. 

\begin{figure}
\centering
\includegraphics[width=0.43\textwidth]{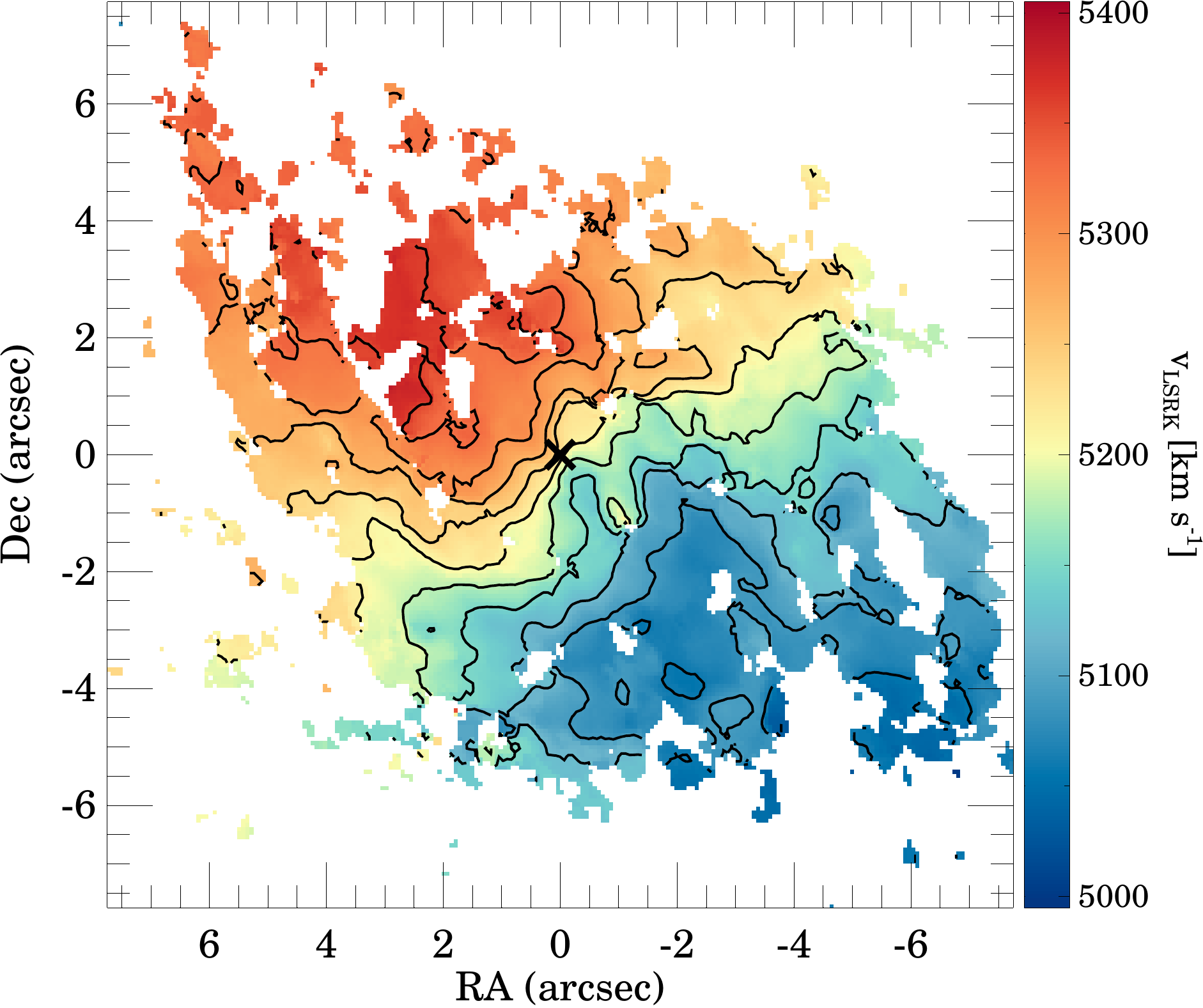}
\caption{CO(2--1) isovelocity contours in steps of 30\,km\,s$^{-1}$ from $v($\textsc{lsrk}$)=$ 5000 to 5400\,km\,s$^{-1}$. The black cross indicates the kinematic center.
\label{fig_isovel}}
\end{figure}

We calculated the position of the dynamical center of the CO(2--1) emission by locating the maximum of the directional derivative of the velocity field (Figure \ref{fig_isovel}; see also \citealt{Arribas1997}). Then, we aligned the peak of the stellar mass distribution traced by the NICMOS and SINFONI near-IR continuum with the CO(2--1) dynamical center.

\section{Analysis and results}\label{s_analysis}

\subsection{Molecular and ionized gas morphology}

\begin{figure}
\centering
\includegraphics[width=0.43\textwidth]{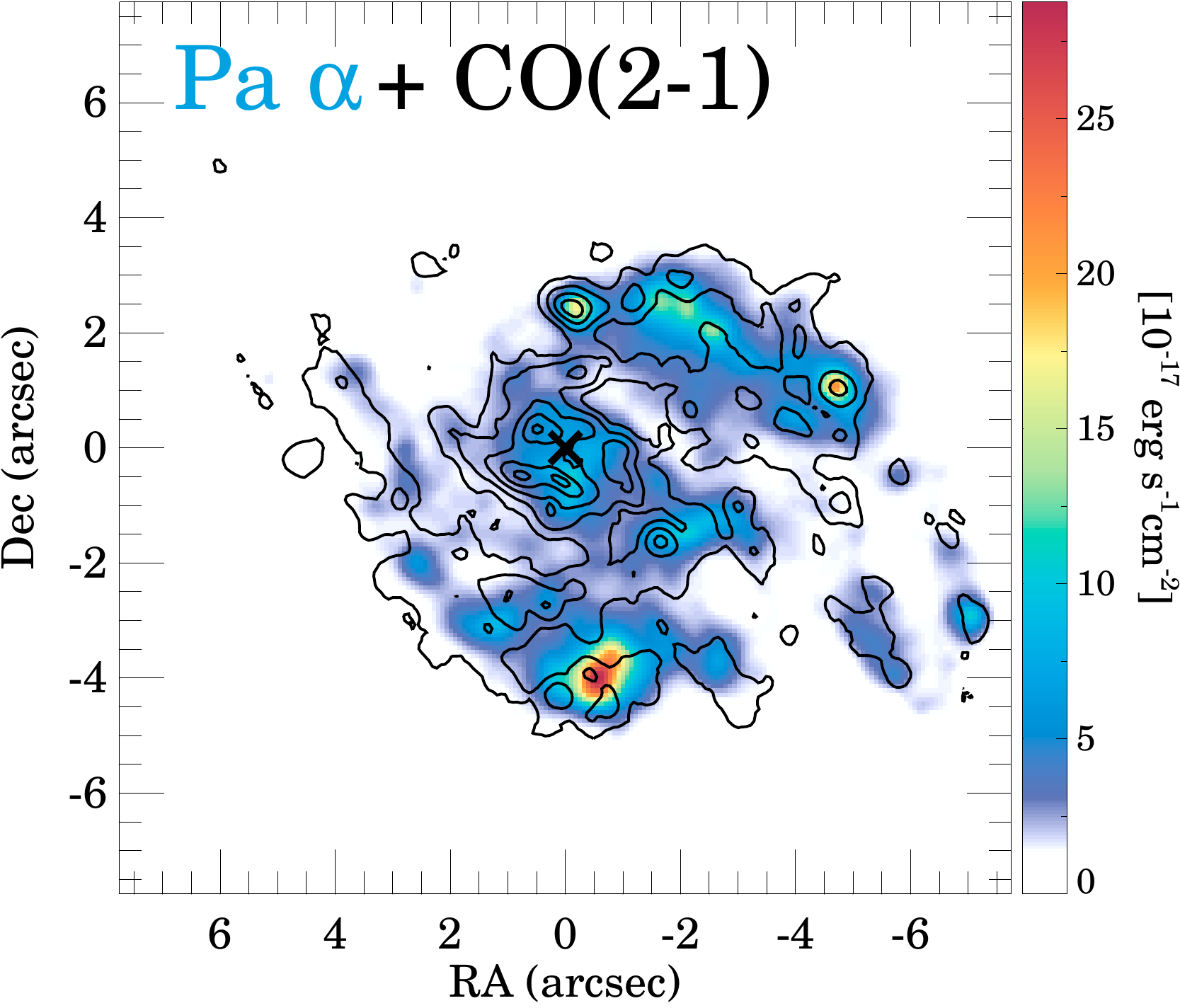}
\caption{CO(2--1) integrated emission contours overlaid on the Pa$\alpha$ emission map of IC~4687. Both maps are shown at the same spatial resolution. The black cross indicates the kinematic center. The contour levels of the CO(2--1) integrated flux correspond to 0.43, 1.1, 1.9, 2.8, and 3.4\,Jy\,km\,s$^{-1}$\,beam$^{-1}$.
\label{fig_paco21}}
\end{figure}

In Figure \ref{fig_maps}, we show the Pa$\alpha$ and CO(2--1) integrated flux and peak intensity maps of IC~4687. Four spiral-like arcs are visible in the molecular gas emission. The two more external arms are also evident in the Pa$\alpha$ map (see also Figure \ref{fig_paco21}). The Pa$\alpha$ emission of the southern arc is dominated by a bright $\sim$1\,kpc in diameter region, while in the northern arm it is spread over $\sim$3\,kpc along the arc.

There is a 1\,kpc diameter ring of molecular gas around the nucleus. This ring is spatially coincident with the brightest, hot H$_2$ emission \citep{Piqueras2012}. This ring is relatively weak in the observed Pa$\alpha$ emission. This is mainly because of the higher extinction of the nuclear ring with respect to the rest of SF regions (see Section \ref{ss_regions}).

Figure \ref{fig_paco21} shows the comparison of the Pa$\alpha$ and CO(2--1) emissions. The general agreement is good. As stated above, both the Pa$\alpha$ and CO(2--1) emissions trace similar spiral arcs and a circumnuclear ring, however, on scales of 100\,pc the CO(2--1) and Pa$\alpha$ emission peaks do not always coincide. There are regions where the Pa$\alpha$ emission is strong and there is no clear CO(2--1) peak associated with the emission, while other regions detected in the CO(2--1) maps do not show Pa$\alpha$ emission.

\subsection{Characterization of the regions}\label{ss_regions}

We used the integrated CO(2--1) emission map to define individual emitting regions. Emission peaks above a 10$\sigma$ level were considered. This conservative $\sigma$ level was chosen to exclude residual side lobes produced by the bright central region.
We applied the same procedure to the Pa$\alpha$ map and then we combined both sets of regions.

\begin{figure}[t]
\centering
\includegraphics[width=0.38\textwidth]{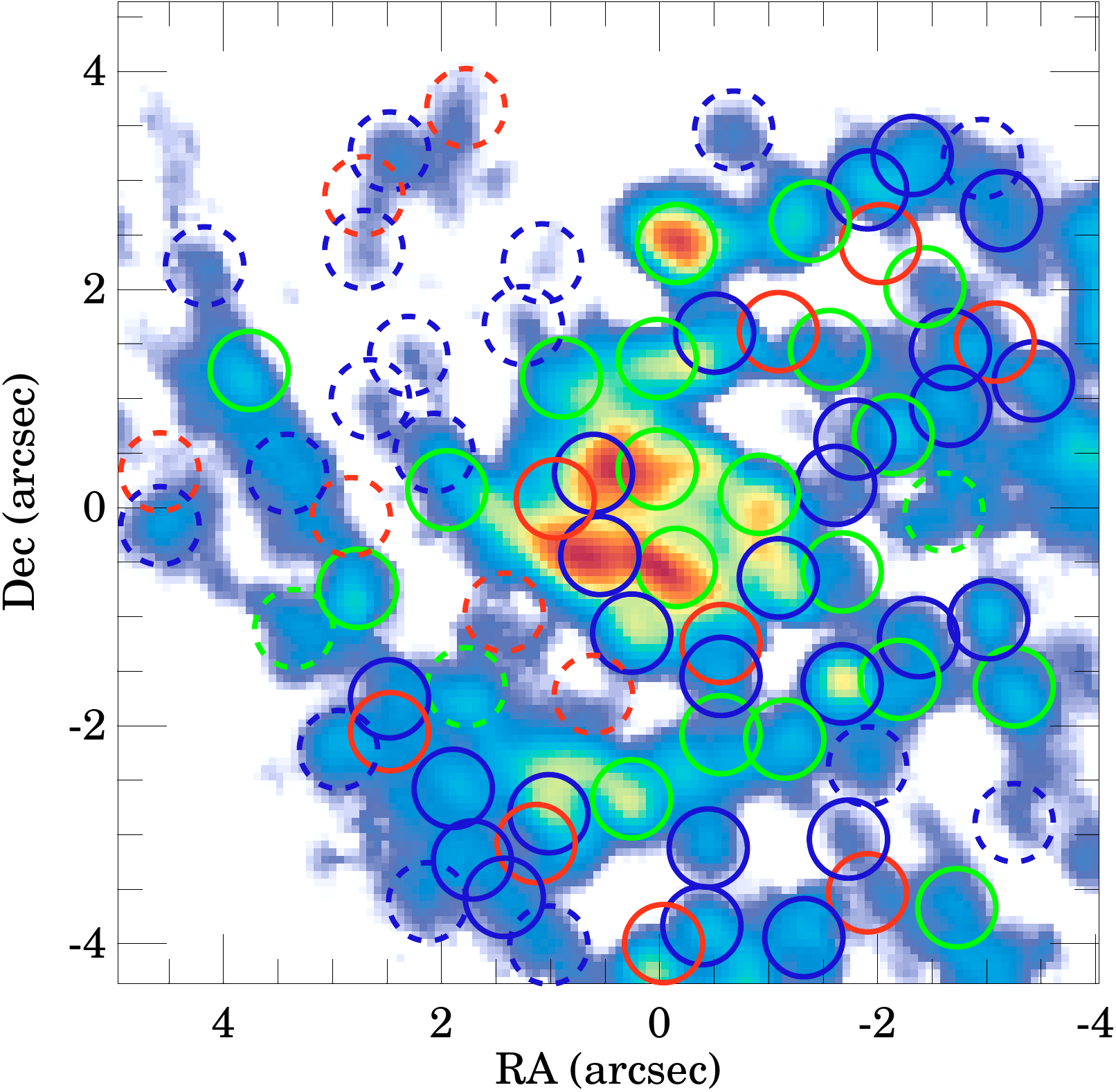}
\caption{Location of the apertures defined for IC~4687  overlaid on the CO(2--1) map. Blue and red circles indicate apertures only detected at $>$10$\sigma$ in the CO(2--1) or the Pa$\alpha$ maps, respectively. Green circles indicate regions with both CO(2--1) and Pa$\alpha$ emission peaks detected at $>$10$\sigma$ level.
Dashed circles indicate regions whose Br$\gamma$ or Br$\delta$ emissions are detected below a 6$\sigma$ level.
\label{fig_regions}}
\end{figure}

In Figure \ref{fig_regions}, we plot the location of these 81 regions.
We assumed that regions with centroids separated by less than 0\farcs35 (ALMA beam) in the CO and Pa$\alpha$ maps correspond to the same physical region. Using this criterion, 23 regions are detected in both the CO(2--1) and Pa$\alpha$ maps. There are 43 and 15 regions detected only in the CO or Pa$\alpha$ maps, respectively.  Both CO(2--1) and Pa$\alpha$ emissions are detected at more than 6$\sigma$ for all these regions, although an emitting clump is seen only in one of the maps.
The diameter of the regions was fixed to 0\farcs7 ($\sim$2 times the ALMA beam), which corresponds to $\sim$250\,pc at the distance of IC~4687. This physical scale is ideal for comparison with previous works (see Section \ref{s_local}).

To measure the CO(2--1) emission of each region, we extracted their spectra and integrated all the channels above a 3$\sigma$ level within the velocity range of the CO(2--1) emission in this object (5000--5400\,km\,s$^{-1}$). 
We estimated the cold molecular gas mass from the CO(2--1) emission using the Galactic CO-to-H$_2$ conversion factor, $\alpha_{\rm CO}^{1-0}=4.35$ (\citealt{Bolatto2013}; see Section \ref{ss_conversion}), and the CO(2--1) to CO(1--0) ratio ($R_{\rm 21}$) of 0.7 derived from the single-dish CO data of this galaxy \citep{Albrecht2007}. This $R_{\rm 21}$ value is similar to that found by \citet{Leroy2013} in nearby spiral galaxies. 
Using this conversion factor, the molecular gas surface density ranges from 10$^{2.3}$ to 10$^{3.4}M_{\odot}$\,pc$^{-2}$ within the 250\,pc of diameter apertures. This corresponds to molecular masses of the individual regions in the range $M_{\rm H_2}=10^7-10^8$\,\Msun, so they likely include several giant molecular clouds.

We estimated the SFR of the regions using the extinction corrected Pa$\alpha$ emission. First, we performed aperture photometry on the Pa$\alpha$ image for each region. To correct the Pa$\alpha$ emission for extinction, we used the Br$\gamma$\slash Br$\delta$ ratio map of \citet{Piqueras2013} and assumed an intrinsic Br$\gamma$\slash Br$\delta$ ratio of 1.52 \citep{Hummer1987} and the \citet[F99]{Fitzpatrick1999} extinction law.
This $A_{\rm K}$ determination is very sensitive to the uncertainty in the Br$\delta$ and Br$\gamma$ fluxes. Therefore, we only considered regions where both the Br$\delta$ and Br$\gamma$ transitions are detected at $>$6$\sigma$. Using this criterion, our final sample includes 54 out of the 81 original regions. Almost 90\%\ of the regions detected in both the CO and Pa$\alpha$ maps fulfill this criterion, while $\sim$40\%\ of the regions detected only in CO or Pa$\alpha$ are excluded. Most of the excluded regions are those at the low-end of the CO and Pa$\alpha$ luminosity distributions. This suggests that we are limited by the sensitivity of the Br$\delta$ and Br$\gamma$ maps, which would be lower than that of the ALMA and HST\slash NICMOS data.

\begin{figure*}[t]
\centering
\hfill
\includegraphics[width=0.45\textwidth]{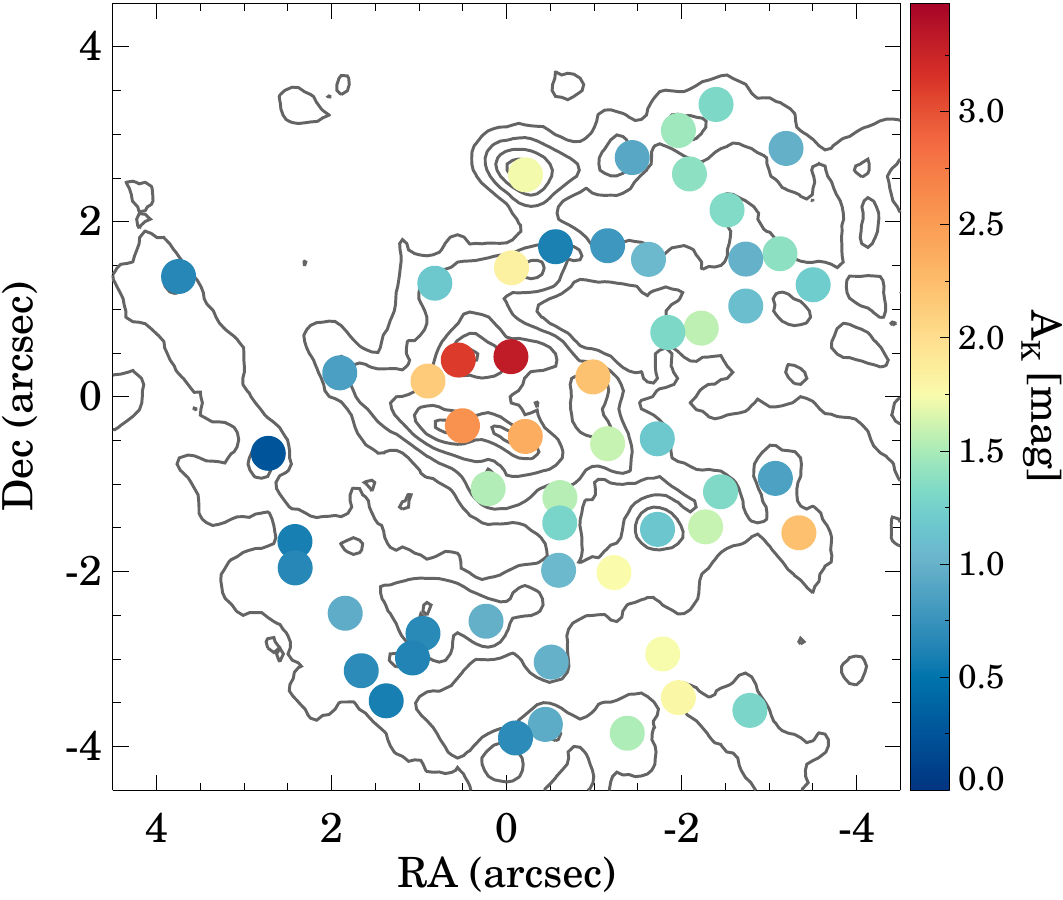}
\hfill
\includegraphics[width=0.45\textwidth]{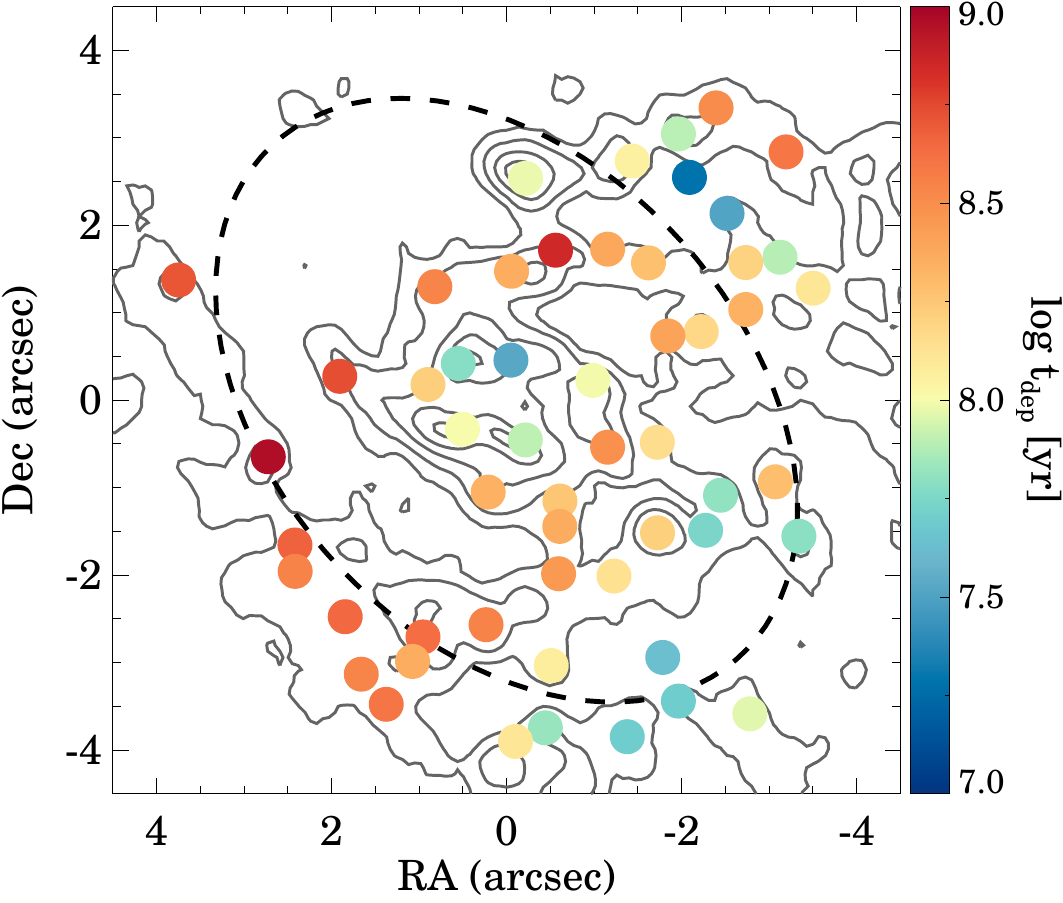}
\hfill
\caption{Spatial distribution of the $A_{\rm K}$ (left) and $t_{\rm dep}$ (right) of the regions. The contours represent the CO(2--1) emission as in Figure \ref{fig_paco21}. The dashed ellipse in the right panel represents the  orbit with a dynamical time of 50\,Myr (see Section \ref{ss_dynamical}).
\label{fig_maps_av_tdep}}
\end{figure*}

The measured extinction range is {$A_{\rm K}$=0.2--3.5\,mag ($A_{\rm V}$=2--30\,mag) with a median $A_{\rm K}$ of 1.3\,mag ($A_{\rm V}$=11\,mag)}. Correcting the observed Pa$\alpha$ emission by this median extinction yields an extinction-corrected flux that is $\sim$4 times the observed flux.
We show the spatial distribution of $A_{\rm K}$ in the left panel of Figure \ref{fig_maps_av_tdep}. The most obscured regions ($A_{\rm K}>1.7$\,mag) are located at the ring of molecular gas around the nucleus, while the regions in the arms have lower $A_{\rm K}$ values.
The extinction-corrected Pa$\alpha$ luminosities of the regions were converted into SFR following the \citet{Kennicutt2012} calibration for H$\alpha$ (assuming H$\alpha$\slash Pa$\alpha$ = 8.58; \citealt{Hummer1987}). For this SFR calibration, \citet{Kennicutt2012} adopted the \citet{Kroupa2001} initial mass function. The SFR surface density in this galaxy is 1--100\,$M_\odot$\,yr$^{-1}$\,kpc$^{-2}$ for the 250\,pc regions.
All of these $\Sigma_{\rm H_2}$ and $\Sigma_{\rm SFR}$ values are multiplied by $\cos 47$\degree\ to correct for the inclination of this galaxy ($i=47$\,\degree; \citealt{Bellocchi2013}).

\begin{figure}
\centering
\includegraphics[width=0.41\textwidth]{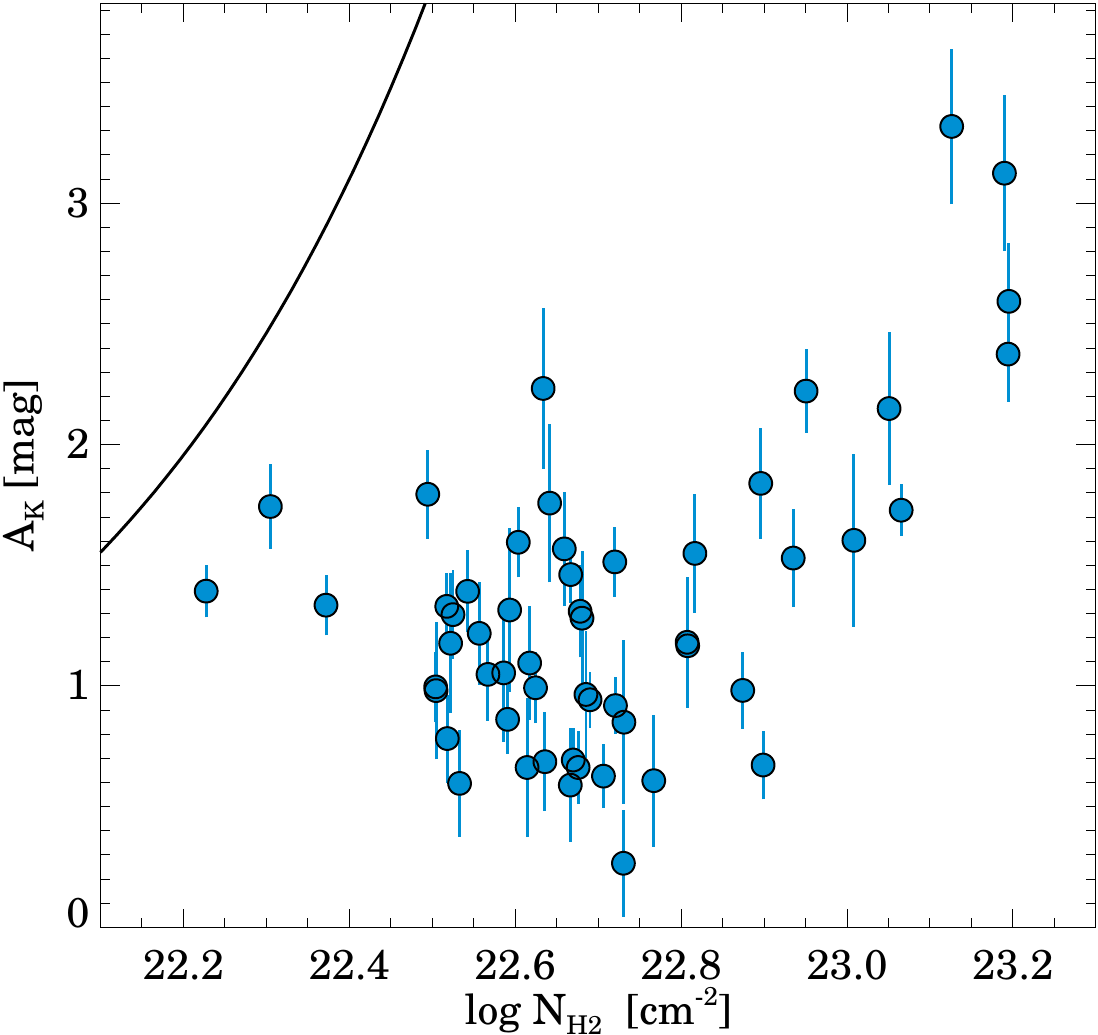}
\caption{Relation between the H$_2$ column density ($N_{\rm H_2}$) and the extinction ($A_{\rm K}$) in IC~4687 for the 250\,pc regions. The solid black line corresponds to { $N_{\rm H_2}$\slash $A_{\rm K}$ = 8.1$\times10^{21}$\,cm$^{-2}$\,mag$^{-1}$ \citep{Bohlin1978}}. \label{fig_av_nh2}}\end{figure}

\subsection{Systematic uncertainties}\label{ss_systematic}

Both the SFR and the cold molecular gas surface density estimates are affected by several systematic effects. These effects have been widely studied in the past (e.g., \citealt{Rahman2011, Liu2011, Schruba2011, Bolatto2013, Genzel2013, Casasola2015}). Therefore, in this section, we briefly discuss the possible systematic effects due to the region selection, SFR tracer, extinction correction, and CO-to-H$_2$ conversion factor we used.

\subsubsection{Region selection}

In the local spiral M~33, for physical scales of 300\,pc, \citet{Schruba2010} found that the depletion time ($t_{\rm dep}$) is shorter by a factor of $\sim$3 for apertures centered on H$\alpha$ peaks than for apertures centered on CO peaks. 
For IC~4687, the average $\log t_{\rm dep}\slash {\rm yr}$ is 8.3$\pm$0.3 and 8.2$\pm$0.4 for the CO and Pa$\alpha$ selected regions, respectively. That is, we do not see any significant difference between the CO and Pa$\alpha$ selected regions with the 250\,pc apertures in this galaxy in terms of $t_{\rm dep}$.
Therefore, in the following, we do not distinguish between CO and Pa$\alpha$ selected regions.

\subsubsection{Extinction correction and SFR tracer}\label{ss_syst_extinct}

We used the Galactic F99 extinction law to correct for dust obscuration effects on IC~4687. This choice is appropriate because we use hydrogen recombination lines (Pa$\alpha$, Br$\delta$, and Br$\gamma$) to derive $A_{\rm K}$ values { ($A_{\rm V}$ = 8.6$\times A_{\rm K}$)}.  We also tried the \citet{Calzetti2000} extinction law. Both laws have similar { A$_{\rm K}$\slash A$_{\rm Pa\alpha}$ ratios ($\sim$1.3), but we find that $A_{\rm K}$ values derived using the \citet{Calzetti2000} law are $\sim$2.2 times lower than following the Galactic law}. This is owing to the steeper slope of this law in the range 1.9--2.2\,\micron\ where the Br$\gamma$ and Br$\delta$ transitions lie. { This difference only affects very obscured regions ($A^{\rm F99}_{\rm K}>2$\,mag) where the SFR derived assuming the \citeauthor{Calzetti2000} law would be $\sim$3 times lower.}

Near-IR transitions yield higher $A_{\rm K}$ { (or $A_{\rm V}$)} values than optical transitions (e.g., Balmer decrement). This is because the relative contribution of highly obscured regions to the total line emission is higher for the near-IR lines. Therefore, { the equivalent $A_{\rm V}$} estimated from near-IR lines is higher as well. 
Consequently, SF laws depend on how the extinction correction is applied. For instance, using optical tracers, the derived SFR varies up to a factor of 10 for extremely obscured galaxies and for the slope of the SF laws (e.g., \citealt{Genzel2013}). In the near-IR, the extinction effects are greatly reduced, so for our case, we estimate that the uncertainties due to the application of the extinction correction are only a factor of $\sim$2 for the $A_{\rm K}$ range of this object.

In addition, extinction corrections can be performed region by region or using an average $A_{\rm K}$ value. For IC~4687, we find that the results are similar on average when using the region-by-region extinction and the integrated extinction (see Section \ref{s_higz}). However, like \citet{Genzel2013}, we find that the relation between the SFR and cold molecular gas is flatter if we use an average extinction value. This is because of the relation between the $A_{\rm K}$ and the H$_2$ column density (Figure \ref{fig_av_nh2}). Regions with more cold molecular gas are more extinguished, so they are undercorrected when we apply the average extinction. Therefore, the relation between the SFR and the cold molecular gas is flatter when the average $A_{\rm K}$ is assumed.

Finally, to check the extinction correction applied, in Figure \ref{fig_av_nh2} we plot the relation between the H$_2$ column density, which is derived from the $\Sigma_{\rm H_2}$ values, and $A_{\rm K}$. In Galactic regions there is a correlation between these two quantities \citep{Bohlin1978, Pineda2010}. In IC~4687, this trend is relatively weak (Spearman's rank correlation coefficient $r_{\rm s}=0.30$, probability of no correlation $p=0.04$), although, as expected, regions with higher H$_2$ column densities tend to have higher $A_{\rm K}$. In particular, this occurs for regions with $A_{\rm K}>1.6$\,mag and $\log N_{\rm H_2} {\rm (cm^{-2})}>22.7$. For lower $A_{\rm K}$ and $N_{\rm H_2}$ values, this relation disappears in IC~4687 at these spatial scales. For comparison, we also plot in Figure \ref{fig_av_nh2} the Galactic relation as a solid line. The measured $A_{\rm K}$ values in IC~4687 are systematically lower than the Galactic prediction by factor of $\sim$6.
{ This suggests that the dust properties and\slash or geometry of the star-forming regions of this object differ from those found in Galactic regions.}
In Section \ref{ss_conversion}, we explain that the Galactic $\alpha_{\rm CO}$ factor is favored for IC~4687. However, we emphasize that using the $\alpha_{\rm CO}$ of ULIRGs, which is 5-7 times lower than the Galactic $\alpha_{\rm
CO}$ factor \citep{Bolatto2013}, would reconcile the observed H$_2$ column densities and $A_{\rm K}$ with the Galactic relation.

\subsubsection{CO-to-H$_2$ conversion factor}\label{ss_conversion}

The derived cold molecular gas masses depend on the $\alpha_{\rm CO}$ conversion factor used.
In Section \ref{ss_regions}, we assumed that the Galactic $\alpha_{\rm CO}$ factor is a good choice for IC~4687. However, we could expect a lower conversion factor, similar to that of ULIRGs, in this galaxy because of its high specific SFR (sSFR = SFR\slash stellar mass), $\sim$0.4\,Gyr$^{-1}$ \citep{Pereira2011}. \citet{Genzel2015} proposed that galaxies with high sSFR, that is, galaxies that lie above the MS of SF galaxies, have reduced $\alpha_{\rm CO}$ factors. IC~4687 has a ${\rm sSFR}\sim6$ times higher than a local MS galaxy with the same stellar mass \citep{Whitaker2012}. Therefore, using the $\alpha_{\rm CO}$ factor of ULIRGs could be justified.

However, it is not clear if the integrated CO-to-H$_2$ conversion factor of ULIRGs, where CO emission is not likely confined to individual molecular clouds \citep{Bolatto2013}, applies to our 250\,pc regions in IC~4687. 
In addition, IC~4687 is not a strongly interacting galaxy or a merger like most local ULIRGs (Figure \ref{fig_maps}); it has a velocity field dominated by rotation (Figure \ref{fig_isovel}), although it is perturbed by noncircular motions. In addition, the morphology of the CO emission of IC~4687 resembles that of a normal spiral galaxy (see \citealt{Leroy2008}) with the SF spread over a region of several kpc.
Therefore, it is possible that the cold molecular gas properties (turbulence, temperature, and density) of IC~4687 differ from those of local ULIRGs where a lower $\alpha_{\rm CO}$ factor is required. 
In fact, in a single-dish survey of local U\slash LIRGs, \citet{Papadopoulos2012b} found that near-Galactic $\alpha_{\rm CO}$ values for U\slash LIRGs are possible when the contribution from high density gas ($n>10^4$\,cm$^{-3}$) is taken into account. Also, in the case of IC~4687, if we used the $\alpha_{\rm CO}$ of ULIRGs, the $t_{\rm dep}$ of the regions would be extremely short, that is, almost 100 times shorter than those of local spiral galaxies (see Section \ref{ss_systematic2}).

\section{Comparison with local galaxies}\label{s_local}

\begin{figure}
\centering
\includegraphics[width=0.42\textwidth]{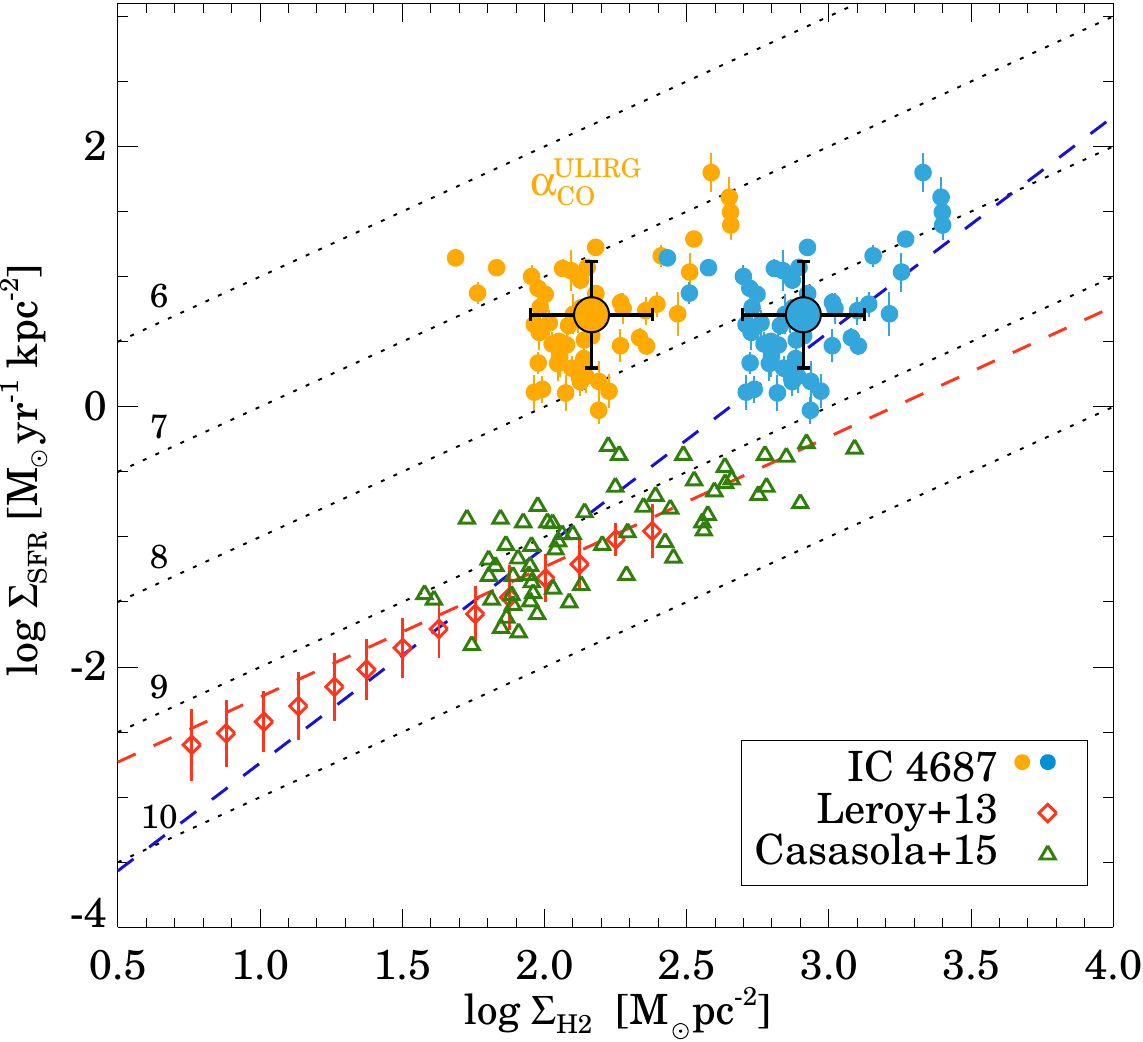}
\caption{SFR surface density as function of the molecular gas surface density. The circles correspond to the 250\,pc regions defined for IC~4687 (Section \ref{ss_regions}). For the orange circles  we applied the CO-to-H$_2$ conversion factor of local ULIRGs, while for the blue circles we applied the Milky Way conversion factor. The large circles indicate the average values for the IC~4687 regions. The green triangles and red diamonds indicate spatially resolved measurements in nearby galaxies by \citet{Casasola2015} and \citet{Leroy2013}, respectively. The molecular gas tracer for all of them is the CO(2--1) transition. The dotted lines mark constant $\log t_{\rm dep}$ times. The dashed blue line is the best fit to all the data ($N$=1.6), and the dashed red line represents the best fit excluding the IC~4687 data ($N$=1.0). \label{fig_sfr_law_local}}
\end{figure}

In Figure \ref{fig_sfr_law_local}, we compare the spatially resolved (200--500\,pc) SFR and molecular gas surface densities of nearby galaxies presented by \citet{Leroy2008} and \citet{Casasola2015} with those of IC~4687. \citet{Leroy2008} studied a sample of 23 nearby ($d<15$\,Mpc) normal spiral galaxies, while \citet{Casasola2015} studied four nearby ($d<20$\,Mpc) low-luminosity AGN.

Figure \ref{fig_sfr_law_local} shows that IC~4687 regions have high molecular gas surface densities, $\log \Sigma_{\rm H_2}{\rm (M_{\odot}\,pc^{-2})} = 2.9\pm0.2$, close to the high end of the $\Sigma_{\rm H_2}$ distribution observed in nearby active galaxies. Moreover, the IC~4687 regions form stars more rapidly than normal galaxies do. These regions have $\log \Sigma_{\rm SFR}{\rm (M_{\odot}\,yr^{-1}\,kpc^{-2})} = 0.7\pm0.4$, which is a factor of $\sim$10 higher than the most extreme values measured in nearby galaxies. This is consistent with the general behavior of local LIRGs, as their \ion{H}{ii} regions are typically a factor of 10 more luminous than those in normal star-forming galaxies \citep{AAH2002, AAH06s}.
Consequently, the $t_{\rm dep}$ of the IC~4687 regions is 160$^{+750}_{-140}$\,Myr (average $\log t_{\rm dep}\slash{\rm yr} = 8.2\pm0.4$). This is approximately one order of magnitude shorter than in nearby galaxies, which is 1--2\,Gyr \citep{Bigiel2008, Bigiel2011, Leroy2013, Casasola2015} for similar physical scales.

\subsection{Systematic uncertainties}\label{ss_systematic2}

In this section, we use data from the local studies of \citet{Leroy2008} and \citet{Casasola2015}. \citet{Leroy2008} used the H$\alpha$+24\micron\ luminosities to derive the SFR, so our results are directly comparable. \citet{Casasola2015} instead used the extinction-corrected H$\alpha$ luminosity. They used the integrated Pa$\alpha$\slash H$\alpha$ ratio to derive this correction, so we expect their SFR to be systematically underestimated when compared to ours (Section \ref{ss_syst_extinct}). Since their galaxies are less extinguished ($A_{\rm K}\sim0.2$\,mag) than the LIRG IC~4687, we estimate that this difference is less than a factor of 2. 

These local studies assume a Galactic CO-to-H$_2$ conversion factor, which is the adopted factor for IC~4687. Consequently, if this factor is valid for IC~4687, all the cold molecular gas surface density comparisons should be consistent. However, in Figure \ref{fig_sfr_law_local}, we also plot the SF law assuming an $\alpha_{\rm CO}$ factor typical of ULIRGs \citep{Downes1998} for IC~4687. Using this factor, the molecular gas masses and depletion times are reduced by a factor of $\sim$5. Therefore, the regions in IC~4687 would have $\Sigma_{\rm H_2}$ similar to those of normal galaxies, but $\Sigma_{\rm SFR}$ $\sim$100 times higher. Actually, if we apply the $\alpha_{\rm CO}$ factor of ULIRGs to IC~4687, this galaxy would be an extreme starburst compared to local and high-$z$ galaxies (Section \ref{s_higz}). In principle, we do not expect such extreme behavior in a weakly interacting spiral galaxy such as IC~4687. Therefore, we consider the Galactic $\alpha_{\rm CO}$ factor preferred for IC~4687.

\subsection{Higher star formation efficiency?}

On average, the SF regions of the LIRG IC~4687 have higher cold molecular gas surface densities than those in other nearby galaxies measured on similar spatial scales when assuming the same $\alpha_{\rm CO}$ (Figure \ref{fig_sfr_law_local}).
 There is some overlap, however, with the regions measured by \citet{Casasola2015} in the range $M_{\rm H_2} = 10^{2.5}-10^{3.1}$\,\Msun\,pc$^{-2}$. 

If the SFR were linearly correlated with the amount of molecular gas (see e.g., \citealt{Bigiel2008}), we would expect higher SFR densities in IC~4687, but also similar depletion times. However, the depletion times in IC~4687 are 10 times shorter than in nearby galaxies. Therefore, for IC~4687, the SFR surface density does not follow the relation observed in nearby galaxies, even in overlapping mass range. 

Alternatively, a nonlinear SF law fits the data with a power-law index of $N=1.6$ (Figure \ref{fig_sfr_law_local}), which is similar to the indexes derived for galaxy integrated data (see Section \ref{s_intro}). However, if we exclude the IC~4687 data from the fit, a linear relation is recovered. That is, the nonlinearity of the relation is only due to the IC~4687 regions. Sub-kpc resolved observations of more extreme starbursts will be needed to determine if they follow a nonlinear SF law or if the SF efficiency (SFE) is actually bimodal since it is more efficient in starburst galaxies (see Section \ref{ss_bimodal}).

\subsection{Dispersion of $t_{\rm dep}$ in IC~4687}

We find that the $t_{\rm dep}$ scatter within IC~4687 regions is relatively high, at 0.4\,dex. A similar, although slightly lower, dispersion of the $t_{\rm dep}$ values is found in nearby galaxies observed at similar spatial resolution \citep{Leroy2013, Casasola2015}. In addition, in IC~4687 (Figure \ref{fig_sfr_law_local}) the correlation between the molecular gas and the SFR surface densities is weak ($r_{\rm S}=0.24$, $p=0.08$). This suggests that, on scales of 250\,pc, the relation between the SFR and the cold molecular gas breaks in this galaxy, or at least, it is hidden by the scatter.

Some works (e.g., \citealt{Onodera2010, Schruba2010, Kruijssen2014}) argue that the time evolution of the SF regions plays a key role in explaining the $t_{\rm dep}$ scatter when high spatial resolution data is used.
Actually, at high spatial resolution (75\,pc) the distributions of the CO and ionized gas emissions are different (e.g., \citealt{Schruba2010}). This is also partially true on scales of 130\,pc for IC~4687 (Figure \ref{fig_paco21}). Therefore, the evolutionary state of the molecular clouds in IC~4687 could give rise to the scatter in the SFR vs. cold molecular gas relation.  With the current observations, however, it is not possible to establish the evolutionary state of the regions in IC~4687, so we cannot test this hypothesis.

Additional scatter is produced by the selected SFR tracers (e.g., \citealt{Schruba2011}). We use the extinction-corrected Pa$\alpha$ emission as a tracer of the SFR. The Pa$\alpha$ traces the ionizing radiation produced by young stars and it is detectable for clusters younger than $\sim$10\,Myr \citep{Kennicutt2012}.
Therefore, our SFR estimates are only sensitive to recent SF ($<$10\,Myr), which might be more variable than the SFR averaged over longer time periods ($\sim$100\,Myr) traced by the UV or IR continuum (e.g., \citealt{Schruba2011, Casasola2015}). This short-term SFR variability can also produce part of the scatter seen in Figure \ref{fig_sfr_law_local}.

Finally, when the mass of the young SF regions is low ($<$10$^5$\,$M_\odot$), the incomplete sampling of the IMF can induce large variations in the SFR tracers (e.g., \citealt{Verley2010}). To test this effect, we estimated the mass of the young stars in each region from the Pa$\alpha$ luminosity. For an instantaneous burst of SF, the \textsc{starburst}\,99 code \citep{Leitherer1999} provides the ionizing radiation produced by a cluster as a function of time. Therefore, assuming that the regions of IC~4687 are close to the peak of the ionizing radiation production (i.e., 0--3\,Myr old), we determine that these young clusters have stellar masses between 10$^{5.5}$ and 10$^{7}$\,\Msun\ (these are lower limits if the regions are older than 3\,Myr; see also \citealt{AAH2002}). \citet{Cervino2002b} showed that for young clusters more massive than 10$^{5}$\,$M_\odot$ (stellar mass) the uncertainties due to the IMF sampling are less than 25\,\%. Consequently, it is not likely that the IMF sampling has any effect on the correlation shown in Figure \ref{fig_sfr_law_local} at the SFR level of IC~4687  using 250\,pc apertures.

\begin{figure}
\centering
\includegraphics[width=0.42\textwidth]{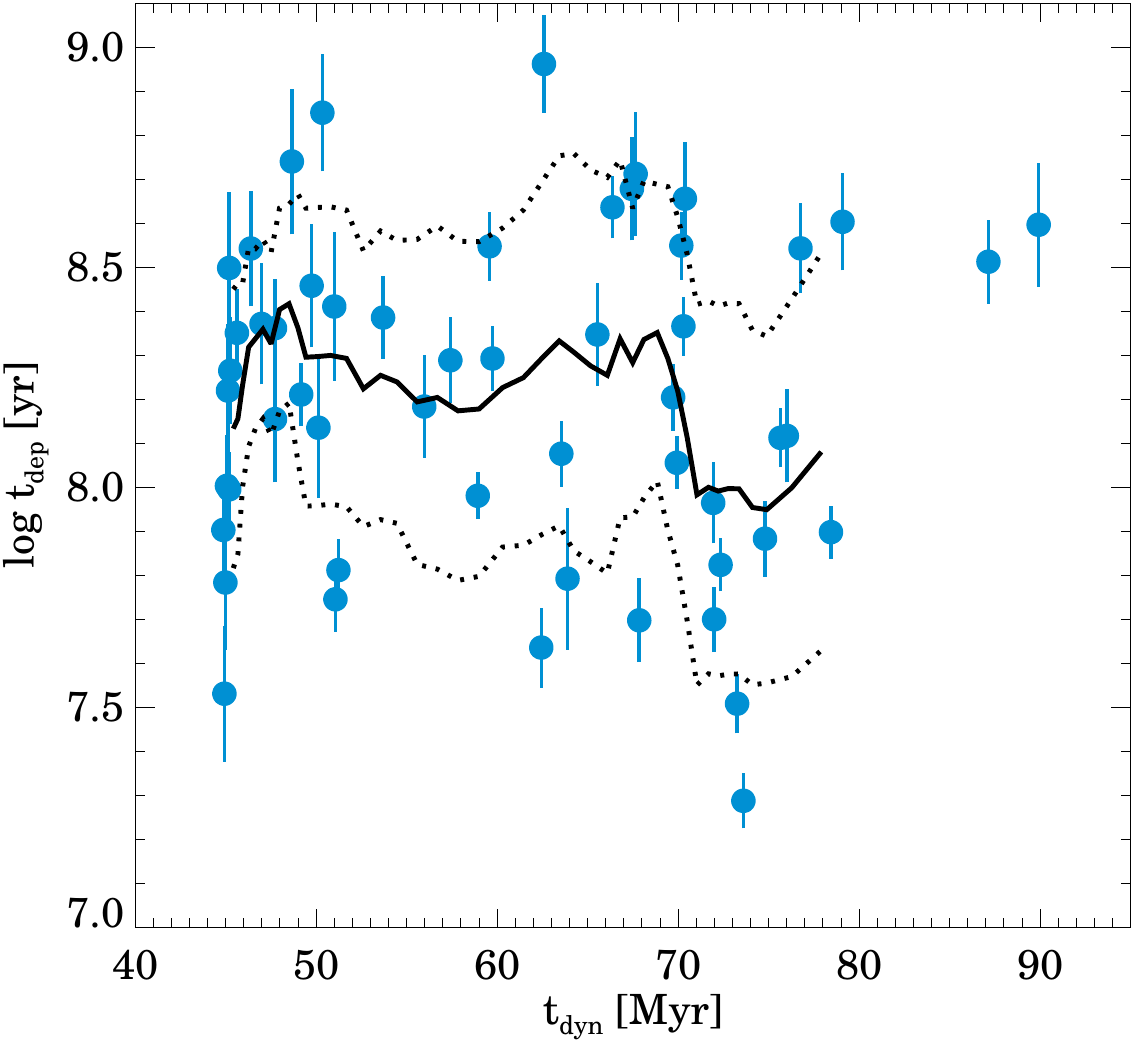}
\caption{Depletion time ($t_{\rm dep}$) as a function of the local dynamical time ($t_{\rm dyn}$). The solid black line indicates the running average $t_{\rm dep}$ value over ten regions. The dotted lines represent the 1$\sigma$ dispersion of this average.\label{fig_tdep}}
\end{figure}

\subsection{Local dynamical time}\label{ss_dynamical}

An alternative formulation of the SF law uses the dynamical time (or orbital time $=$ 2$\pi r$\slash $v_{\rm rot}$) to normalize the molecular gas surface density (e.g., \citealt{Kennicutt2007}).  This formulation uses an average dynamical time for integrated measurements; it is able to recover a universal SF law valid for objects with high SFE, such as ULIRGs or sub-mm galaxies, and for normal spirals (see also Section \ref{ss_bimodal}).

For our resolved observations, it is possible to estimate the dynamical time of each region from their deprojected radius (assuming $i=47\degree$ and a major axis PA of 39$\degree$) and the rotation curve derived by \citet{Bellocchi2013} using kinemetry \citep{Krajnovic2006}. In Figure \ref{fig_tdep}, we show that the depletion time (or SFE) does not depend on the dynamical time ($r_{\rm S}=0.02$, $p=0.86$). In the right panel of Figure \ref{fig_maps_av_tdep}, we show the spatial distribution of depletion times where no clear trends are seen. This absence of correlation is also seen in resolved observations of normal spirals \citep{Leroy2008}. Therefore, the local dynamical time does not seem to influence the local SFE at spatial scales of $\sim$250\,pc.

\begin{figure}
\centering
\includegraphics[width=0.42\textwidth]{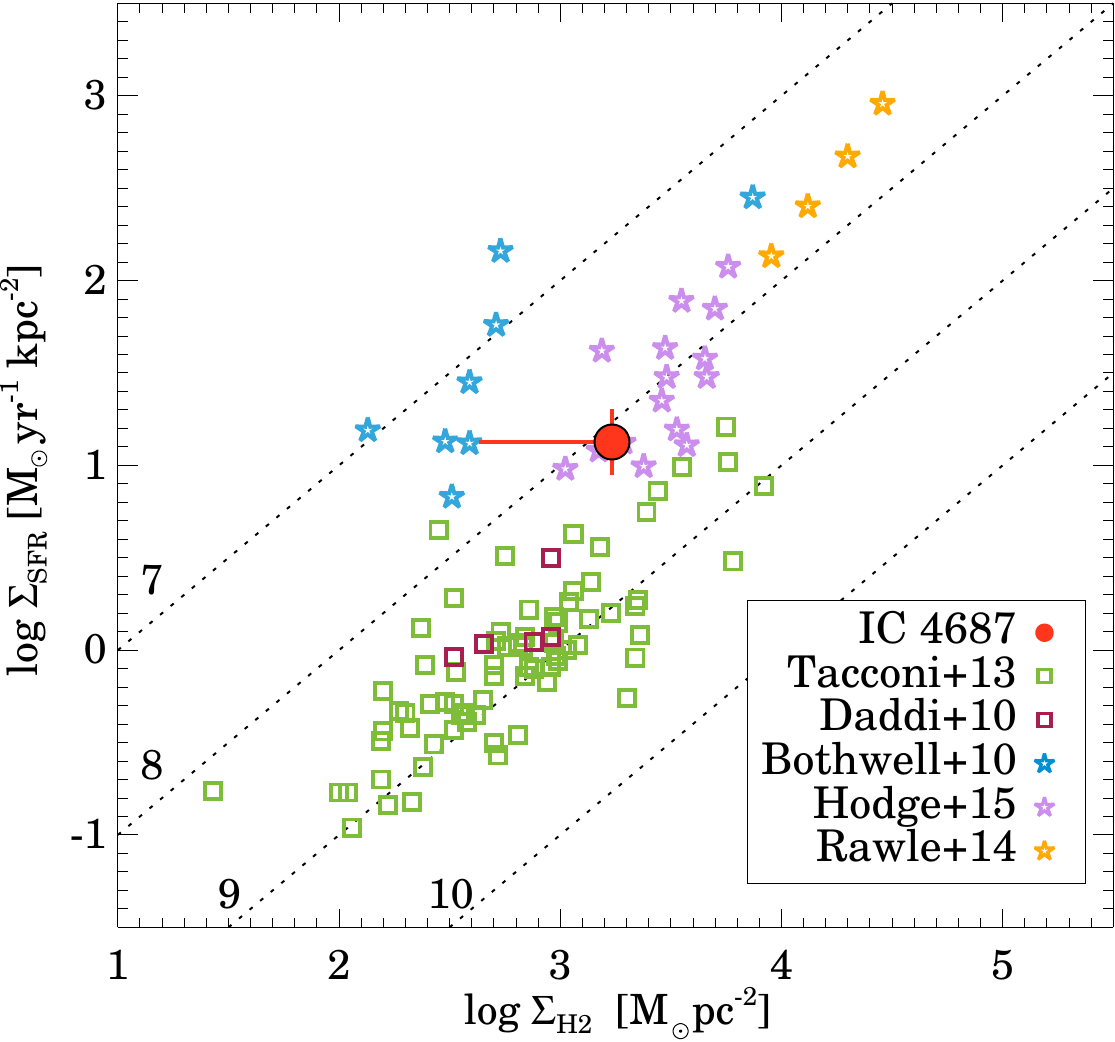}
\caption{Comparison of the SFR surface density as a function of the molecular gas surface density for high-$z$ MS and sub-mm galaxies and IC~4687. The red circle indicates the integrated measurement for IC~4687, as described in Section \ref{s_higz}, using the Galactic $\alpha_{\rm CO}$ factor. The error bars indicate systematic uncertainties due to the extinction correction in $\Sigma_{\rm SFR}$ (vertical) and a change in the $\alpha_{\rm CO}$ factor from Galactic (assumed) to that considered for ULIRGs (horizontal; see Section \ref{ss_systematic}).
The green and red squares correspond to $z\sim 1.2$ and 2.2 MS SF galaxies from \citet{Tacconi2013} and BzK $z\sim 1.5$ galaxies from \citet{Daddi2010b}. For both datasets we used the Galactic $\alpha_{\rm CO}$ factor. The blue, purple, and orange stars correspond to sub-millimeter galaxies at $z\sim2$, 4.0, and 5.2 from \citet{Bothwell2010}, \citet{Hodge2015}, and \citet{Rawle2014}, respectively, for which a ULIRG-like $\alpha_{\rm CO}$ factor was applied. The dotted lines indicate constant $\log t_{\rm dep}$ times. \label{fig_sfr_law_highz}}
\end{figure}

\section{Comparison with high-$z$ galaxies}\label{s_higz}

\subsection{Integrated properties of IC~4687}\label{ss_integrated}

The SF laws derived for high-$z$ galaxies are mostly based on integrated measurements. Therefore, it is useful to calculate the integrated properties of IC~4687 using an approach comparable to high-$z$ studies.

We limited our integrated study to the 3$\times$3\,kpc$^2$ area covered by the field of view of SINFONI (see Section \ref{s_data}) to obtain an accurate measurement of the extinction. This area contains about 85\%\ of the total CO(2--1) and 90\%\ of the observed Pa$\alpha$ emissions. Therefore, by limiting our analysis to this area, we only miss 10--15\%\ of the total emission.

We derived an integrated extinction { of $A_{\rm K}=1.2\pm 0.1$\,mag ($A_{\rm V}=10.1\pm 0.8$\,mag)} based on the integrated Br$\gamma$\slash Br$\delta$ ratio. The total SFR, 43$\pm$4\,\Msun\,yr$^{-1}$, is calculated from the extinction-corrected Br$\gamma$ integrated flux\footnote{This Br$\gamma$ flux is multiplied by 12.1 \citep{Storey1995} to obtain the Pa$\alpha$ equivalent flux.} because the \textit{HST}\slash NICMOS Pa$\alpha$ image is not sensitive to diffuse Pa$\alpha$ emission due to its
small pixel size; this diffuse emission would be included in integrated measurements of high-$z$ objects. This SFR is $\sim$1.5 times higher than that derived from the IR luminosity, but this difference is within the assumed systematic uncertainties (see Section \ref{ss_systematic}). From the integrated CO(2--1) emission, we derive a total cold molecular gas mass of 5.5$\times$10$^9$\,\Msun.

In IC~4687, the extinction-corrected Pa$\alpha$ emission from the regions defined in Section \ref{ss_regions} accounts for $\sim$70\%\ of the integrated SFR derived here. 
For the cold molecular gas, $\sim$65\%\ of the CO(2--1) emission comes from these regions. These two fractions are similar. Therefore, the integrated $t_{\rm dep}$ agrees with the average $t_{\rm dep}$ of the individual regions.

The effective CO(2--1) emitting area (that containing 50\%\ of the total CO(2--1) emission) has a $R_{\rm 1/2}$ of 1.0\,kpc, which corresponds to an $\sim$30\%\ lower area than that of the individual regions combined. Therefore, because of the increased integrated CO(2--1) and Pa$\alpha$ fluxes from diffuse emission and the lower emitting area estimate, both the integrated SFR and cold molecular gas surface densities are $\sim$2 times higher than the average values of the resolved regions in IC~4687, although the depletion times are similar.

\subsection{High-$z$ counterparts of IC~4687}

In Figure \ref{fig_sfr_law_highz}, we compare the integrated SF law of IC~4687 with those obtained for MS $z\sim$1.2--2.2 galaxies \citep{Tacconi2013}, $z\sim$1.5 BzK galaxies \citep{Daddi2010b}, and submillimeter galaxies \citep{Bothwell2010, Rawle2014, Hodge2015}. 

We find that the integrated H$_2$ and SFR surface densities of IC~4687 lie at the high end of the distributions of values measured for high-$z$ MS galaxies. The $\Sigma_{\rm SFR}$ and $\Sigma_{\rm H_2}$ values directly depends on the size of the emitting region, which is not easy to estimate for integrated galaxies (see Section \ref{ss_integrated} and \citealt{Arribas2012}). The depletion time, however, is independent of the emitting region size. Therefore, we focus on the $t_{\rm dep}$ differences in this section.

A galaxy with the sSFR of IC~4687 (sSFR = $\sim$0.4\,Gyr$^{-1}$) would be a MS galaxy at $z\sim0.9$ (see Section \ref{ss_conversion}). As shown in Figure \ref{fig_sfr_law_highz}, high-$z$ MS galaxies, in general, are less efficient than IC~4687 forming stars. On average, they have $t_{\rm dep}$ that are 6 times longer than IC~4687 (using the Galactic $\alpha_{\rm CO}$ for IC~4687). Therefore, even if these $z=1.2-2.2$ galaxies have sSFR, SFR, and stellar masses similar to IC~4687, they are more similar to local starbursts in terms of $t_{\rm dep}$ \citep{Tacconi2013}.
This can be explained by the correlation between the $t_{\rm dep}$ and the sSFR normalized by the sSFR of a MS galaxy between $z=0$ and 3 \citep{Genzel2015}. Since, the sSFR of IC~4687 is $\sim$6 times higher than the local MS sSFR (Section \ref{ss_conversion}), we expect a shorter depletion time in this local LIRG than in MS galaxies, at least for $z<3$.

On the other hand, the $t_{\rm dep}$ of IC~4687 is similar to that of high-$z$ sub-mm galaxies (\citealt{Rawle2014, Hodge2015}; Figure \ref{fig_sfr_law_highz}). The amount of cold molecular gas in IC~4687 and these sub-mm galaxies is also similar ($\sim$10$^{10}$\,\Msun), however, this depends on the CO-to-H$_2$ conversion used. For the sub-mm galaxies, the $\alpha_{\rm CO}$ used is similar to that of local ULIRGs, that is, it is lower than the Galactic $\alpha_{\rm CO}$ used for IC~4687. Consequently, if we would apply the ULIRG $\alpha_{\rm CO}$ factor to IC~4687, its $t_{\rm dep}$ would be at the high end of the $t_{\rm dep}$ range measured in high-$z$ sub-mm galaxies (Figure \ref{fig_sfr_law_highz}).

\subsection{Systematic uncertainties}

For high-$z$ galaxies, the SFR is mainly obtained from the spectral energy distribution fitting. This kind of  analysis includes the IR emission, therefore it is possible that our SFR derived from the Pa$\alpha$ luminosity are underestimated by a factor of two (see \citealt{Piqueras2015}). The $\alpha_{\rm CO}$ factor applied to high-$z$ galaxies depends on the object class (Galactic factor for MS galaxies; ULIRG-like factor for sub-mm galaxies). Therefore, the comparison with the molecular gas surface density of IC~4687 is somewhat uncertain. For reference, in Figure \ref{fig_sfr_law_highz} we represent the range of $\Sigma_{\rm H_2}$ assuming the Galactic and  ULIRG $\alpha_{\rm CO}$ factors.

\subsection{Bimodal SF law}\label{ss_bimodal}

Some studies have shown that the SF laws have a bimodal behavior with a factor 3--4 lower $t_{\rm dep}$ in local U\slash LIRGs than in normal SF galaxies (e.g., \citealt{Daddi2010b, Genzel2010, GarciaBurillo2012}). We observe this bimodal behavior when we compare MS galaxies (Figures \ref{fig_sfr_law_local} and \ref{fig_sfr_law_highz}) with IC~4687.

To recover a universal SF law for integrated measurements of galaxies, several alternative formulations are proposed. We discuss two of them here. The first  is to normalize the cold molecular gas surface density by the dynamical time ($\Sigma_{\rm H_2}\slash t_{\rm dyn}$; \citealt{Silk1997, Tan2000}). When this normalization is applied to integrated measurements, a global $t_{\rm dyn}$ is used (e.g., \citealt{Kennicutt1998b, Daddi2010b}). In this case, the global dynamical times of U\slash LIRGs ($\sim$45\,Myr) are 4--5 times shorter than those of spirals ($\sim$370\,Myr), so the lower $t_{\rm dep}$ values of U\slash LIRGs are compensated and a universal relation between $\Sigma_{\rm SFR}$ and $\Sigma_{\rm H_2}\slash t_{\rm dyn}$ is obtained (e.g., \citealt{Genzel2010, GarciaBurillo2012}). The dynamical time of the outer regions of IC~4687 is $\sim$80\,Myr. Consequently, IC~4687 would lie in this universal relation as well (cf. equation 21 in \citealt{GarciaBurillo2012}). 

In Section \ref{ss_dynamical}, we showed that the SFE does not depend on the local $t_{\rm dyn}$ in IC~4687. Thus, SF does not seem to be strongly affected by the local effects of the disk rotation. Therefore, the normalization by the global $t_{\rm dyn}$ might be a simplification of the physical mechanisms leading to this universal SF law relation.

Alternatively, the free-fall time is another proposed normalization for the cold molecular gas surface density ($\Sigma_{\rm H_2}\slash t_{\rm ff}$; \citealt{Krumholz2005}) to recover a universal SF law. The $t_{\rm ff}$ is proportional to 1\slash$\rho^{0.5}$, where $\rho$ is the molecular gas volume density \citep{Binney1987}. Therefore, systems with higher $\rho$ have lower $t_{\rm ff}$. If the molecular gas density is higher in U\slash LIRGs than in normal spirals \citep{Gao2004}, the $t_{\rm ff}$ normalization would recover a universal relation as well. 

\section{Conclusions}\label{s_conclusions}

We have analyzed the resolved (250\,pc scales) and integrated SF law in the local LIRG IC~4687. 
This is one of the first studies of the SF laws on a starburst dominated LIRG at these spatial scales.
We combined new interferometric ALMA CO(2--1) observations with existing \textit{HST}\slash NICMOS Pa$\alpha$ narrowband imaging and VLT\slash SINFONI near-IR integral field spectroscopy to obtain accurate cold molecular gas masses and extinction-corrected SFR estimates.
The main conclusions of our analysis are the following:
\begin{enumerate}
\item We defined 54 regions with a diameter of 250\,pc centered at the CO and Pa$\alpha$ emission peaks. 
The resolved $\Sigma_{\rm H_2}$ values of IC~4687 lie at the high end of the values observed in local galaxies at these spatial resolutions. Whereas the $\Sigma_{\rm SFR}$ are almost a factor of 10 higher than those of local galaxies for similar $\Sigma_{\rm H_2}$. 
For the resolved regions of IC~4687, the correlation between $\Sigma_{\rm H_2}$ and $\Sigma_{\rm SFR}$ is weak ($r_{\rm S} = 0.25$, $p=0.08$). This suggests that the SF law breaks downs in this galaxy on scales of 250\,pc.

\item Compared with resolved SF laws in local galaxies, IC~4687 forms stars more efficiently. The range of $t_{\rm dep}$ of the individual regions is 20--900\,Myr with an average of 160\,Myr. This is almost one order of magnitude shorter than that of local galaxies. For these estimates, we used a Galactic $\alpha_{\rm CO}$ conversion factor; using an ULIRG-like factor would make the $t_{\rm dep}$ even shorter by an additional factor of 4--5.

\item The 1$\sigma$ scatter in the $t_{\rm dep}$ values is 0.4\,dex. We suggest that this can be due to the rapid time evolution of the SFR tracer we used (Pa$\alpha$). We rule out that the IMF sampling causes the observed scatter for this galaxy because of the high young stellar masses (10$^{5.5-7}$\,\Msun) of the studied regions. We also show that the local dynamical time does not significantly affect the SF efficiency in IC~4687 (up to $\sim$1.5\,kpc away from the nucleus).

\item The galaxy integrated $\log\,\Sigma_{\rm H_2}{(M_{\odot}\rm \,pc^{-2})}=2.6-3.2$ and $\log\,\Sigma_{\rm SFR}{(M_{\odot}\rm \,yr^{-1}\,kpc^{-2})}=1.1\pm0.2$ of IC~4687 make this object have a $t_{\rm dep}$ $\sim$6 times shorter than MS high-$z$ galaxies. 
The $\Sigma_{\rm H_2}$ lies at the high end of the $\Sigma_{\rm H_2}$ distribution of high-$z$ MS objects, whereas the $\Sigma_{\rm SFR}$ is $\sim$10 times higher than in high-$z$ objects with similar $\Sigma_{\rm H_2}$. There are some high-$z$ MS galaxies with comparable $\Sigma_{\rm SFR}$ levels, although they have higher $\Sigma_{\rm H_2}$ values than IC~4687.

\item Our results suggest that SF is more efficient in IC~4687 than in nearby  star-forming galaxies. This agrees with some works that propose the existence of a bimodal SF law. After normalizing the $\Sigma_{\rm H_2}$ by the {global} dynamical time, IC~4687 lies in the universal SF law. However, since the {local} dynamical time does not affect the local SFE, this {global} dynamical time normalization could be contrived. Alternatively, a normalization using the $t_{\rm ff}$ might recover a universal SF law. 
The $t_{\rm ff}$ depends on the volume density, therefore, future high spatial resolution observations of dense molecular gas in LIRGs and normal galaxies will reveal whether the local $t_{\rm ff}$ has any influence on the SFE at sub-kpc scales.
\end{enumerate}

\begin{acknowledgements}
{ We thank the referee, Erik Rosolowsky, for his comments, which helped improve this paper.}
We acknowledge support from the Spanish Plan Nacional de Astronom\'ia y Astrof\'isica through grants AYA2010-21161-C02-01, AYA2012-32295, AYA2012-39408-C2-1, AYA2012-31447, and FIS2012-32096.
This paper makes use of the following ALMA data: ADS/JAO.ALMA\#2013.1.00271.S. ALMA is a partnership of ESO (representing its member states), NSF (USA) and NINS (Japan), together with NRC (Canada) and NSC and ASIAA (Taiwan), in cooperation with the Republic of Chile. The Joint ALMA Observatory is operated by ESO, AUI/NRAO and NAOJ.
Partially based on observations collected at the European Organisation for Astronomical Research in the Southern Hemisphere, Chile, programmes 077.B-0151A.

\end{acknowledgements}


\begin{thebibliography}{64}
\expandafter\ifx\csname natexlab\endcsname\relax\def\natexlab#1{#1}\fi

\bibitem[{{Albrecht} {et~al.}(2007){Albrecht}, {Kr{\"u}gel}, \&
  {Chini}}]{Albrecht2007}
{Albrecht}, M., {Kr{\"u}gel}, E., \& {Chini}, R. 2007, \aap, 462, 575

\bibitem[{{Alonso-Herrero} {et~al.}(2012){Alonso-Herrero}, {Pereira-Santaella},
  {Rieke}, \& {Rigopoulou}}]{AAH2012a}
{Alonso-Herrero}, A., {Pereira-Santaella}, M., {Rieke}, G.~H., \& {Rigopoulou},
  D. 2012, \apj, 744, 2

\bibitem[{{Alonso-Herrero} {et~al.}(2006){Alonso-Herrero}, {Rieke}, {Rieke},
  {Colina}, {P{\'e}rez-Gonz{\'a}lez}, \& {Ryder}}]{AAH06s}
{Alonso-Herrero}, A., {Rieke}, G.~H., {Rieke}, M.~J., {et~al.} 2006, \apj, 650,
  835

\bibitem[{{Alonso-Herrero} {et~al.}(2002){Alonso-Herrero}, {Rieke}, {Rieke}, \&
  {Scoville}}]{AAH2002}
{Alonso-Herrero}, A., {Rieke}, G.~H., {Rieke}, M.~J., \& {Scoville}, N.~Z.
  2002, \aj, 124, 166

\bibitem[{{Arribas} {et~al.}(2012){Arribas}, {Colina}, {Alonso-Herrero},
  {Rosales-Ortega}, {Monreal-Ibero}, {Garc{\'{\i}}a-Mar{\'{\i}}n},
  {Garc{\'{\i}}a-Burillo}, \& {Rodr{\'{\i}}guez-Zaur{\'{\i}}n}}]{Arribas2012}
{Arribas}, S., {Colina}, L., {Alonso-Herrero}, A., {et~al.} 2012, \aap, 541,
  A20

\bibitem[{{Arribas} {et~al.}(2014){Arribas}, {Colina}, {Bellocchi}, {Maiolino},
  \& {Villar-Mart{\'{\i}}n}}]{Arribas2014}
{Arribas}, S., {Colina}, L., {Bellocchi}, E., {Maiolino}, R., \&
  {Villar-Mart{\'{\i}}n}, M. 2014, \aap, 568, A14

\bibitem[{{Arribas} {et~al.}(1997){Arribas}, {Mediavilla},
  {Garc{\'{\i}}a-Lorenzo}, \& {del Burgo}}]{Arribas1997}
{Arribas}, S., {Mediavilla}, E., {Garc{\'{\i}}a-Lorenzo}, B., \& {del Burgo},
  C. 1997, \apj, 490, 227

\bibitem[{{Bellocchi} {et~al.}(2013){Bellocchi}, {Arribas}, {Colina}, \&
  {Miralles-Caballero}}]{Bellocchi2013}
{Bellocchi}, E., {Arribas}, S., {Colina}, L., \& {Miralles-Caballero}, D. 2013,
  \aap, 557, A59

\bibitem[{{Bigiel} {et~al.}(2008){Bigiel}, {Leroy}, {Walter}, {Brinks}, {de
  Blok}, {Madore}, \& {Thornley}}]{Bigiel2008}
{Bigiel}, F., {Leroy}, A., {Walter}, F., {et~al.} 2008, \aj, 136, 2846

\bibitem[{{Bigiel} {et~al.}(2011){Bigiel}, {Leroy}, {Walter}, {Brinks}, {de
  Blok}, {Kramer}, {Rix}, {Schruba}, {Schuster}, {Usero}, \&
  {Wiesemeyer}}]{Bigiel2011}
{Bigiel}, F., {Leroy}, A.~K., {Walter}, F., {et~al.} 2011, \apjl, 730, L13

\bibitem[{{Binney} \& {Tremaine}(1987)}]{Binney1987}
{Binney}, J. \& {Tremaine}, S. 1987, {Galactic dynamics}

\bibitem[{{Bohlin} {et~al.}(1978){Bohlin}, {Savage}, \& {Drake}}]{Bohlin1978}
{Bohlin}, R.~C., {Savage}, B.~D., \& {Drake}, J.~F. 1978, \apj, 224, 132

\bibitem[{{Bolatto} {et~al.}(2013){Bolatto}, {Wolfire}, \&
  {Leroy}}]{Bolatto2013}
{Bolatto}, A.~D., {Wolfire}, M., \& {Leroy}, A.~K. 2013, \araa, 51, 207

\bibitem[{{Bothwell} {et~al.}(2010){Bothwell}, {Chapman}, {Tacconi}, {Smail},
  {Ivison}, {Casey}, {Bertoldi}, {Beswick}, {Biggs}, {Blain}, {Cox}, {Genzel},
  {Greve}, {Kennicutt}, {Muxlow}, {Neri}, \& {Omont}}]{Bothwell2010}
{Bothwell}, M.~S., {Chapman}, S.~C., {Tacconi}, L., {et~al.} 2010, \mnras, 405,
  219

\bibitem[{{Briggs}(1995)}]{Briggs1995PhDT}
{Briggs}, D.~S. 1995, PhD thesis, New Mexico Institute of Mining and Technology

\bibitem[{{Calzetti} {et~al.}(2000){Calzetti}, {Armus}, {Bohlin}, {Kinney},
  {Koornneef}, \& {Storchi-Bergmann}}]{Calzetti2000}
{Calzetti}, D., {Armus}, L., {Bohlin}, R.~C., {et~al.} 2000, \apj, 533, 682

\bibitem[{{Casasola} {et~al.}(2015){Casasola}, {Hunt}, {Combes}, \&
  {Garc{\'{\i}}a-Burillo}}]{Casasola2015}
{Casasola}, V., {Hunt}, L., {Combes}, F., \& {Garc{\'{\i}}a-Burillo}, S. 2015,
  \aap, 577, A135

\bibitem[{{Cazzoli} {et~al.}(2015, submitted){Cazzoli}, {Arribas}, \&
  et~al.}]{Cazzoli2015}
{Cazzoli}, S., {Arribas}, S., \& et~al. 2015, submitted

\bibitem[{{Cervi{\~n}o} {et~al.}(2002){Cervi{\~n}o}, {Valls-Gabaud},
  {Luridiana}, \& {Mas-Hesse}}]{Cervino2002b}
{Cervi{\~n}o}, M., {Valls-Gabaud}, D., {Luridiana}, V., \& {Mas-Hesse}, J.~M.
  2002, \aap, 381, 51

\bibitem[{{Daddi} {et~al.}(2010){Daddi}, {Elbaz}, {Walter}, {Bournaud},
  {Salmi}, {Carilli}, {Dannerbauer}, {Dickinson}, {Monaco}, \&
  {Riechers}}]{Daddi2010b}
{Daddi}, E., {Elbaz}, D., {Walter}, F., {et~al.} 2010, \apjl, 714, L118

\bibitem[{{Downes} \& {Solomon}(1998)}]{Downes1998}
{Downes}, D. \& {Solomon}, P.~M. 1998, \apj, 507, 615

\bibitem[{{Fitzpatrick}(1999)}]{Fitzpatrick1999}
{Fitzpatrick}, E.~L. 1999, \pasp, 111, 63

\bibitem[{{Gao} \& {Solomon}(2004)}]{Gao2004}
{Gao}, Y. \& {Solomon}, P.~M. 2004, \apj, 606, 271

\bibitem[{{Garc{\'{\i}}a-Burillo} {et~al.}(2012){Garc{\'{\i}}a-Burillo},
  {Usero}, {Alonso-Herrero}, {Graci{\'a}-Carpio}, {Pereira-Santaella},
  {Colina}, {Planesas}, \& {Arribas}}]{GarciaBurillo2012}
{Garc{\'{\i}}a-Burillo}, S., {Usero}, A., {Alonso-Herrero}, A., {et~al.} 2012,
  \aap, 539, A8

\bibitem[{{Genzel} {et~al.}(2010){Genzel}, {Tacconi}, {Gracia-Carpio},
  {Sternberg}, {Cooper}, {Shapiro}, {Bolatto}, {Bouch{\'e}}, {Bournaud},
  {Burkert}, {Combes}, {Comerford}, {Cox}, {Davis}, {Schreiber},
  {Garcia-Burillo}, {Lutz}, {Naab}, {Neri}, {Omont}, {Shapley}, \&
  {Weiner}}]{Genzel2010}
{Genzel}, R., {Tacconi}, L.~J., {Gracia-Carpio}, J., {et~al.} 2010, \mnras,
  407, 2091

\bibitem[{{Genzel} {et~al.}(2013){Genzel}, {Tacconi}, {Kurk}, {Wuyts},
  {Combes}, {Freundlich}, {Bolatto}, {Cooper}, {Neri}, {Nordon}, {Bournaud},
  {Burkert}, {Comerford}, {Cox}, {Davis}, {F{\"o}rster Schreiber},
  {Garc{\'{\i}}a-Burillo}, {Gracia-Carpio}, {Lutz}, {Naab}, {Newman},
  {Saintonge}, {Shapiro Griffin}, {Shapley}, {Sternberg}, \&
  {Weiner}}]{Genzel2013}
{Genzel}, R., {Tacconi}, L.~J., {Kurk}, J., {et~al.} 2013, \apj, 773, 68

\bibitem[{{Genzel} {et~al.}(2015){Genzel}, {Tacconi}, {Lutz}, {Saintonge},
  {Berta}, {Magnelli}, {Combes}, {Garc{\'{\i}}a-Burillo}, {Neri}, {Bolatto},
  {Contini}, {Lilly}, {Boissier}, {Boone}, {Bouch{\'e}}, {Bournaud}, {Burkert},
  {Carollo}, {Colina}, {Cooper}, {Cox}, {Feruglio}, {F{\"o}rster Schreiber},
  {Freundlich}, {Gracia-Carpio}, {Juneau}, {Kovac}, {Lippa}, {Naab}, {Salome},
  {Renzini}, {Sternberg}, {Walter}, {Weiner}, {Weiss}, \& {Wuyts}}]{Genzel2015}
{Genzel}, R., {Tacconi}, L.~J., {Lutz}, D., {et~al.} 2015, \apj, 800, 20

\bibitem[{{Hodge} {et~al.}(2015){Hodge}, {Riechers}, {Decarli}, {Walter},  {Carilli}, {Daddi}, \& {Dannerbauer}}]{Hodge2015}
{Hodge}, J.~A., {Riechers}, D., {Decarli}, R., {et~al.} 2015, \apjl, 798, L18

\bibitem[{{Hummer} \& {Storey}(1987)}]{Hummer1987}
{Hummer}, D.~G. \& {Storey}, P.~J. 1987, \mnras, 224, 801

\bibitem[{{Kennicutt} \& {Evans}(2012)}]{Kennicutt2012}
{Kennicutt}, R.~C. \& {Evans}, N.~J. 2012, \araa, 50, 531

\bibitem[{{Kennicutt}(1998)}]{Kennicutt1998b}
{Kennicutt}, Jr., R.~C. 1998, \apj, 498, 541

\bibitem[{{Kennicutt} {et~al.}(2007){Kennicutt}, {Calzetti}, {Walter}, {Helou},
  {Hollenbach}, {Armus}, {Bendo}, {Dale}, {Draine}, {Engelbracht}, {Gordon},
  {Prescott}, {Regan}, {Thornley}, {Bot}, {Brinks}, {de Blok}, {de Mello},
  {Meyer}, {Moustakas}, {Murphy}, {Sheth}, \& {Smith}}]{Kennicutt2007}
{Kennicutt}, Jr., R.~C., {Calzetti}, D., {Walter}, F., {et~al.} 2007, \apj,
  671, 333

\bibitem[{{Krajnovi{\'c}} {et~al.}(2006){Krajnovi{\'c}}, {Cappellari}, {de
  Zeeuw}, \& {Copin}}]{Krajnovic2006}
{Krajnovi{\'c}}, D., {Cappellari}, M., {de Zeeuw}, P.~T., \& {Copin}, Y. 2006,
  \mnras, 366, 787

\bibitem[{{Kroupa}(2001)}]{Kroupa2001}
{Kroupa}, P. 2001, \mnras, 322, 231

\bibitem[{{Kruijssen} \& {Longmore}(2014)}]{Kruijssen2014}
{Kruijssen}, J.~M.~D. \& {Longmore}, S.~N. 2014, \mnras, 439, 3239

\bibitem[{{Krumholz} \& {McKee}(2005)}]{Krumholz2005}
{Krumholz}, M.~R. \& {McKee}, C.~F. 2005, \apj, 630, 250

\bibitem[{{Leitherer} {et~al.}(1999){Leitherer}, {Schaerer}, {Goldader},
  {Delgado}, {Robert}, {Kune}, {de Mello}, {Devost}, \&
  {Heckman}}]{Leitherer1999}
{Leitherer}, C., {Schaerer}, D., {Goldader}, J.~D., {et~al.} 1999, \apjs, 123,
  3

\bibitem[{{Leroy} {et~al.}(2008){Leroy}, {Walter}, {Brinks}, {Bigiel}, {de
  Blok}, {Madore}, \& {Thornley}}]{Leroy2008}
{Leroy}, A.~K., {Walter}, F., {Brinks}, E., {et~al.} 2008, \aj, 136, 2782\bibitem[{{Leroy} {et~al.}(2013){Leroy}, {Walter}, {Sandstrom}, {Schruba},
  {Munoz-Mateos}, {Bigiel}, {Bolatto}, {Brinks}, {de Blok}, {Meidt}, {Rix},
  {Rosolowsky}, {Schinnerer}, {Schuster}, \& {Usero}}]{Leroy2013}
{Leroy}, A.~K., {Walter}, F., {Sandstrom}, K., {et~al.} 2013, \aj, 146, 19

\bibitem[{{Liu} {et~al.}(2011){Liu}, {Koda}, {Calzetti}, {Fukuhara}, \&
  {Momose}}]{Liu2011}
{Liu}, G., {Koda}, J., {Calzetti}, D., {Fukuhara}, M., \& {Momose}, R. 2011,
  \apj, 735, 63

\bibitem[{{McKee} \& {Ostriker}(2007)}]{McKee2007}
{McKee}, C.~F. \& {Ostriker}, E.~C. 2007, \araa, 45, 565

\bibitem[{{McMullin} {et~al.}(2007){McMullin}, {Waters}, {Schiebel}, {Young},
  \& {Golap}}]{McMullin2007}
{McMullin}, J.~P., {Waters}, B., {Schiebel}, D., {Young}, W., \& {Golap}, K.
  2007, in Astronomical Society of the Pacific Conference Series, Vol. 376,
  Astronomical Data Analysis Software and Systems XVI, ed. R.~A. {Shaw},
  F.~{Hill}, \& D.~J. {Bell}, 127

\bibitem[{{Onodera} {et~al.}(2010){Onodera}, {Kuno}, {Tosaki}, {Kohno},
  {Nakanishi}, {Sawada}, {Muraoka}, {Komugi}, {Miura}, {Kaneko}, {Hirota}, \&
  {Kawabe}}]{Onodera2010}
{Onodera}, S., {Kuno}, N., {Tosaki}, T., {et~al.} 2010, \apjl, 722, L127

\bibitem[{{Papadopoulos} {et~al.}(2012){Papadopoulos}, {van der Werf},
  {Xilouris}, {Isaak}, \& {Gao}}]{Papadopoulos2012b}
{Papadopoulos}, P.~P., {van der Werf}, P., {Xilouris}, E., {Isaak}, K.~G., \&
  {Gao}, Y. 2012, \apj, 751, 10

\bibitem[{{Pereira-Santaella} {et~al.}(2011){Pereira-Santaella},
  {Alonso-Herrero}, {Santos-Lleo}, {Colina}, {Jim{\'e}nez-Bail{\'o}n},
  {Longinotti}, {Rieke}, {Ward}, \& {Esquej}}]{Pereira2011}
{Pereira-Santaella}, M., {Alonso-Herrero}, A., {Santos-Lleo}, M., {et~al.}
  2011, \aap, 535, A93

\bibitem[{{Pineda} {et~al.}(2010){Pineda}, {Goldsmith}, {Chapman}, {Snell},
  {Li}, {Cambr{\'e}sy}, \& {Brunt}}]{Pineda2010}
{Pineda}, J.~L., {Goldsmith}, P.~F., {Chapman}, N., {et~al.} 2010, \apj, 721,
  686

\bibitem[{{Piqueras L{\'o}pez} {et~al.}(2013){Piqueras L{\'o}pez}, {Colina},
  {Arribas}, \& {Alonso-Herrero}}]{Piqueras2013}
{Piqueras L{\'o}pez}, J., {Colina}, L., {Arribas}, S., \& {Alonso-Herrero}, A.
  2013, \aap, 553, A85

\bibitem[{{Piqueras L{\'o}pez} {et~al.}(2012){Piqueras L{\'o}pez}, {Colina},
  {Arribas}, {Alonso-Herrero}, \& {Bedregal}}]{Piqueras2012}
{Piqueras L{\'o}pez}, J., {Colina}, L., {Arribas}, S., {Alonso-Herrero}, A., \&
  {Bedregal}, A.~G. 2012, \aap, 546, A64

\bibitem[{{Piqueras L{\'o}pez} {et~al.}(2015, submitted){Piqueras L{\'o}pez},
  {Colina}, \& et~al.}]{Piqueras2015}
{Piqueras L{\'o}pez}, J., {Colina}, L., \& et~al. 2015, submitted

\bibitem[{{Rahman} {et~al.}(2011){Rahman}, {Bolatto}, {Wong}, {Leroy},
  {Walter}, {Rosolowsky}, {West}, {Bigiel}, {Ott}, {Xue}, {Herrera-Camus},
  {Jameson}, {Blitz}, \& {Vogel}}]{Rahman2011}
{Rahman}, N., {Bolatto}, A.~D., {Wong}, T., {et~al.} 2011, \apj, 730, 72

\bibitem[{{Rahman} {et~al.}(2012){Rahman}, {Bolatto}, {Xue}, {Wong}, {Leroy},
  {Walter}, {Bigiel}, {Rosolowsky}, {Fisher}, {Vogel}, {Blitz}, {West}, \&
  {Ott}}]{Rahman2012}
{Rahman}, N., {Bolatto}, A.~D., {Xue}, R., {et~al.} 2012, \apj, 745, 183

\bibitem[{{Rawle} {et~al.}(2014){Rawle}, {Egami}, {Bussmann}, {Gurwell},
  {Ivison}, {Boone}, {Combes}, {Danielson}, {Rex}, {Richard}, {Smail},
  {Swinbank}, {Altieri}, {Blain}, {Clement}, {Dessauges-Zavadsky}, {Edge},
  {Fazio}, {Jones}, {Kneib}, {Omont}, {P{\'e}rez-Gonz{\'a}lez}, {Schaerer},
  {Valtchanov}, {van der Werf}, {Walth}, {Zamojski}, \& {Zemcov}}]{Rawle2014}
{Rawle}, T.~D., {Egami}, E., {Bussmann}, R.~S., {et~al.} 2014, \apj, 783, 59

\bibitem[{{Rodr{\'{\i}}guez-Zaur{\'{\i}}n}
  {et~al.}(2011){Rodr{\'{\i}}guez-Zaur{\'{\i}}n}, {Arribas}, {Monreal-Ibero},
  {Colina}, {Alonso-Herrero}, \& {Alfonso-Garz{\'o}n}}]{RodriguezZaurin2011}
{Rodr{\'{\i}}guez-Zaur{\'{\i}}n}, J., {Arribas}, S., {Monreal-Ibero}, A.,  {et~al.} 2011, \aap, 527, A60+

\bibitem[{{Schmidt}(1959)}]{Schmidt1959}
{Schmidt}, M. 1959, \apj, 129, 243

\bibitem[{{Schruba} {et~al.}(2011){Schruba}, {Leroy}, {Walter}, {Bigiel},  {Brinks}, {de Blok}, {Dumas}, {Kramer}, {Rosolowsky}, {Sandstrom},
  {Schuster}, {Usero}, {Weiss}, \& {Wiesemeyer}}]{Schruba2011}
{Schruba}, A., {Leroy}, A.~K., {Walter}, F., {et~al.} 2011, \aj, 142, 37

\bibitem[{{Schruba} {et~al.}(2010){Schruba}, {Leroy}, {Walter}, {Sandstrom}, \&
  {Rosolowsky}}]{Schruba2010}
{Schruba}, A., {Leroy}, A.~K., {Walter}, F., {Sandstrom}, K., \& {Rosolowsky},
  E. 2010, \apj, 722, 1699

\bibitem[{{Silk}(1997)}]{Silk1997}
{Silk}, J. 1997, \apj, 481, 703

\bibitem[{{Storey} \& {Hummer}(1995)}]{Storey1995}
{Storey}, P.~J. \& {Hummer}, D.~G. 1995, \mnras, 272, 41

\bibitem[{{Tacconi} {et~al.}(2013){Tacconi}, {Neri}, {Genzel}, {Combes},
  {Bolatto}, {Cooper}, {Wuyts}, {Bournaud}, {Burkert}, {Comerford}, {Cox},
  {Davis}, {F{\"o}rster Schreiber}, {Garc{\'{\i}}a-Burillo}, {Gracia-Carpio},
  {Lutz}, {Naab}, {Newman}, {Omont}, {Saintonge}, {Shapiro Griffin}, {Shapley},
  {Sternberg}, \& {Weiner}}]{Tacconi2013}
{Tacconi}, L.~J., {Neri}, R., {Genzel}, R., {et~al.} 2013, \apj, 768, 74

\bibitem[{{Tan}(2000)}]{Tan2000}
{Tan}, J.~C. 2000, \apj, 536, 173

\bibitem[{{Verley} {et~al.}(2010){Verley}, {Corbelli}, {Giovanardi}, \&
  {Hunt}}]{Verley2010}
{Verley}, S., {Corbelli}, E., {Giovanardi}, C., \& {Hunt}, L.~K. 2010, \aap,
  510, A64

\bibitem[{{Viaene} {et~al.}(2014){Viaene}, {Fritz}, {Baes}, {Bendo},
  {Blommaert}, {Boquien}, {Boselli}, {Ciesla}, {Cortese}, {De Looze}, {Gear},
  {Gentile}, {Hughes}, {Jarrett}, {Karczewski}, {Smith}, {Spinoglio}, {Tamm},
  {Tempel}, {Thilker}, \& {Verstappen}}]{Viaene2014}
{Viaene}, S., {Fritz}, J., {Baes}, M., {et~al.} 2014, \aap, 567, A71

\bibitem[{{Whitaker} {et~al.}(2012){Whitaker}, {van Dokkum}, {Brammer}, \&
  {Franx}}]{Whitaker2012}
{Whitaker}, K.~E., {van Dokkum}, P.~G., {Brammer}, G., \& {Franx}, M. 2012,
  \apjl, 754, L29

\bibitem[{{Yao} {et~al.}(2003){Yao}, {Seaquist}, {Kuno}, \& {Dunne}}]{Yao2003}
{Yao}, L., {Seaquist}, E.~R., {Kuno}, N., \& {Dunne}, L. 2003, \apj, 588, 771

\end{thebibliography}
\end{document}